\documentclass[11pt]{article}

\usepackage {a4}

\usepackage [utf8]{inputenc}
\usepackage [english]{babel}
\usepackage{enumerate}
\usepackage{fullpage}
\usepackage{bbm}
\usepackage{physics}

\usepackage{graphicx}
\usepackage{amsmath,amsthm,amssymb,amsfonts,setspace}
\usepackage{bm}
\usepackage{mathtools}
\usepackage{braket}
\usepackage{float}
\usepackage{color}
\usepackage[usenames,dvipsnames]{xcolor}
\usepackage{hyperref}  
\usepackage[capitalize,nameinlink,noabbrev]{cleveref}
\usepackage{wrapfig}
\usepackage{tikz}
\usepackage[left=1in,right=1in,top=1in,bottom=1in]{geometry}

\hypersetup{colorlinks=true,urlcolor=[rgb]{0,0,0.5},citecolor=[rgb]{0.5,0,0},linkcolor=[rgb]{0,0,0.4}}

\DeclareUnicodeCharacter{00A0}{ }

\newcommand{\id}{\mathbbm{1}}
\newcommand{\Var}{\mathrm{Var}}
\newcommand{\scalar}[2]{\left\langle #1, #2\right\rangle}

\newcommand{\R}{\mathbb{R}}

\newcommand{\C}{\mathbb{C}}
\newcommand{\Pp}{\mathbb{P}}
\newcommand{\E}{\mathbb{E}}

\newcommand{\Z}{\mathbb{Z}}

\newtheorem{theorem}{Theorem}[section]

\newtheorem{lemma}[theorem]{Lemma}
\newtheorem{proposition}[theorem]{Proposition}

\theoremstyle{definition}
\newtheorem{remark}{Remark}[section]

\usepackage[utf8]{inputenc}
\usepackage{todonotes}

\newcommand{\unitary}[1]{ \mathrm{U}(#1)}
\newcommand{\gateset}{ \mathcal{G}}
\newcommand{\states}[1]{ S^{\leq #1}}

\newcommand{\gwords}[1]{ \mathcal{G}^{\leq #1}}

\renewcommand{\H}{\mathcal{H}} 

\newcommand{\GUE}{\mathrm{GUE}}
\newcommand{\HS}{\mathrm{HS}}
\newcommand{\vep}{\varepsilon}

\newcommand{\h}[1]{\boldsymbol{#1}}
\newcommand{\U}{\h{U}} %
\newcommand{\V}{\h{V}} %
\newcommand{\I}{\h{I}} 

\newcommand{\UU}{\mathcal{U}} 
\renewcommand{\S}{\mathcal{S}} 
\newcommand{\G}{\mathcal{G}} 

\usepackage{xcolor}

\newtheorem{Definition}{Definition}

\newtheorem{result}{Result}

\newcommand{\ddiamond}{\mathrm{D}_{\diamond} }
\renewcommand{\d}{\mathrm{D}}

\newcommand{\diaggauss}[1]{\mathrm{DiagGauss(#1)}}
\newcommand{\unitarygauss}[1]{\mathrm{UGauss(#1)}}

\newcommand{\dist}{\mathrm{dist}}

\newcommand{\dhsproj}{\mathrm{D}_\mathrm{HS}} 
\newcommand{\tesc}{t_{\mathrm{escape}}}
\newcommand{\tlar}{t_{\mathrm{jump}}}

\def\cred{\color{red}}
\def\blk{\color{black}}

\usepackage{doi}

\usepackage[backend=biber, maxbibnames=99, style=alphabetic, doi=true, url=false, eprint=false]{biblatex} 

\usetikzlibrary{decorations.pathmorphing}

\usepackage{subcaption}
\usepackage{authblk}

\title{Extremal jumps of circuit complexity of unitary evolutions generated by random Hamiltonians}

\author[1]{Marcin Kotowski \thanks{\!mkotowski@cft.edu.pl}}
\author[1]{Michał Oszmaniec\thanks{\!oszmaniec@cft.edu.pl}}
\author[2]{Michał Horodecki \thanks{\!michal.horodecki@ug.edu.pl}}
\affil[1]{Center for Theoretical Physics, Polish Academy of Sciences,\\ Al.\ Lotnik\'ow 32/46, 02-668
Warszawa, Poland}
\affil[2]{International Centre for Theory of Quantum Technologies, University of Gdansk, Poland}
\date{}                     
\begin{document}

\maketitle

\begin{abstract}
    We investigate circuit complexity of unitaries generated by time evolution of randomly chosen strongly interacting Hamiltonians in finite dimensional Hilbert spaces. Specifically, we focus on two ensembles of random generators -- the so called Gaussian Unitary Ensemble ($\GUE$) and the ensemble of diagonal Gaussian matrices conjugated by Haar random unitary transformations. In both scenarios we prove that the complexity of $\exp(-it H)$ exhibits the following behaviour -- with high probability it reaches the maximal allowed value on the same time scale as needed to escape the neighborhood of the identity consisting of unitaries with trivial (zero) complexity. We furthermore observe similar behaviour for quantum states originating from time evolutions generated by above ensembles and for diagonal unitaries generated from the ensemble of diagonal Gaussian Hamiltonians.  To establish these results we rely  heavily on structural properties of the above ensembles (such as unitary invariance) and concentration of measure techniques.  This gives us a much finer control over the time evolution of complexity compared to techniques previously employed in this context: high-degree moments and frame potentials. 
\end{abstract}


\section{Introduction}

Complexity is a central notion of computer science and quantum complexity is therefore at the heart of quantum computing. Quantum complexity of a unitary operation is the minimal number of some elementary gates that are needed to approximate the unitary to some a priori given accuracy.
The famous Shor's factoring algorithm can be thought as a unitary that enables factoring large numbers, but has relatively small complexity. A typical Haar random unitary is actually the opposite -- a simple counting argument shows that a randomly selected unitary has almost maximal complexity. 
There is a long standing conjecture due to Brown and Susskind \cite{BrownSusskind17} that complexity of unitary evolution $U_t=\exp(-i H t)$ generated by chaotic many-body Hamiltonians  (such as the SYK model \cite{sykcomplexity}) grows linearly with time until exponential time and after doubly exponential time  it undergoes recurrence (i.e. drops down to zero). 

One may ask about the rationale for studying long-time properties of quantum evolutions, given that the success of quantum computing lies in efficient, carefully designed quantum algorithms. One reason is that Brown-Susskind conjecture, if true, implies that a quantum computer cannot efficiently simulate (via fast-forwarding) long time evolution of states and unitaries generated  by generic quantum many-body evolutions.   
Another  prominent reason for which actually the Brown-Susskind conjecture was formulated and then studied for quite some time is its relation to the so called AdS-CFT (Anti-de Sitter - Conformal Field Theory) correspondence.  This conjectured correspondence relates classical gravity in the bulk with quantum field theory on the boundary. The so called wormhole growth paradox has been identified  \cite{SScomp14} (see also \cite{SusskindCCBH14}), namely wormhole (i.e. a sort of "bridge" joining two parts of Penrose diagram for black whole in Anti-de Sitter universe) in the bulk grows for a long time, while, equilibration of the quantum field  on boundary is quite fast 
(logarithmic in black hole entropy). It would thus seem that the slow growth of the wormhole does not have a counterpart in 
the field dynamical behaviour -- there, after logarithmic time, nothing in principle evolves. The idea of \cite{SScomp14} was, then, that there is still some quantity that can grow for a long time -- complexity. Therefore, since the evolution of quantum fields at the boundary is believed to be mimicked by chaotic quantum systems, the paradox could be resolved if such chaotic evolution would exhibit linear growth of complexity. 

For unitary time evolutions  \emph{upper bounds} on the complexity that grow linearly with time can be obtained from \emph{Hamiltonian simulation} algorithms \cite{Lloyd1996,SparseSimulation2015,Low2019hamiltonian} , which work in a multitude of settings including local and sparse Hamiltonians. However, proving unconditional lower bounds is much more challenging. In a pioneering work, Nielsen and collaborators \cite{NielsenCompGeo1,NielsenCompGeo2} established a connection between complexity and geodesic distance in a well-defined sub-Riemannian geometry. Unfortunately, this approach cannot be directly employed to large systems due to the challenging optimization problem of finding the shortest geodesic. Nevertheless, Nielsen's geometric method was applied to compute complexity in free fermionic and bosonic theories in Refs.\cite{FermCompl2018,BosComplexity2018} (see also Ref.\cite{Eisert2021} which utilized this perspective to connect exact complexity based on the entangling power of quantum gates). Additionally, complexity-theoretic conjectures (such as PSPACE $\neq$ BQP) can be used to demonstrate that the complexity of specific Hamiltonian evolutions must grow faster than polynomial for exponentially long times, as shown in Refs.~\cite{aaronson2016complexity,FastForwarding2017,susskind2018black,bohdanowicz2017universal}. Finally, we note that recently in \cite{haystack} the authors give explicit constructions of unitaries (corresponding to  evolutions generated by Hamiltonians with specific algebraic structure) which provably have exponential circuit complexity .

Due to immense difficulty of the Brown-Susskind conjecture in its original formulation, simplified models of evolution were therefore considered. One idea was to analyse evolution generated by random quantum circuits. These models were examined in 
\blk
\cite{complexitygrowth2019,exactcomplexity2021,OHH}. In \cite{complexitygrowth2019} 
growth of complexity was confirmed, though still sublinear. In \cite{exactcomplexity2021} (see also \cite{LiExact2022} for a simplified proof) linear growth was shown, but the notion of complexity was not physical -- the authors used exact complexity and unfortunately their methods were so tightly related to that notion that it is not possible to extend it to cover approximate complexity. Then in \cite{OHH} the physically sound notion of complexity was 
used and it  was shown that after exponential time, complexity becomes maximal, and moreover recurrence after double exponential times was showed.

Another possible simplification is to consider unitary evolutions $\exp(-itH)$ generated by random Hamiltonians $H$.  However, the problem is nontrivial already for the most well studied model of strongly interacting quantum systems -- the so-called Gaussian Unitary Ensemble (GUE) \cite{RMT_anderson_guionnet_zeitoun}. Equilibration properties of such system were examined in 
\cite{Brandao4U,Masanes4U,Vinayak4U,Cramer4U} and compared with 
properties of many-body chaotic systems in \cite{ChaosRMT}. One prominent difference is extremely fast scrambling, which for GUE happens at constant time. Furthermore \cite{ChaosRMT} attempted to lower bound complexity of GUE unitaries  by using findings of earlier work \cite{ChaosDesign} that showed how theory of $k$-designs and machinery of \emph{frame potentials} can be used to this purpose (see also \cite{oliviero} for recent work on GUE and related frame potentials). By approximately and somewhat  heuristically analysing formulas for  high degree frame potentials of the ensemble $\lbrace\exp(-it H)\rbrace_{H\sim\GUE}$ the paper claimed  quadratic ($\sim c t^2$) lower bound on complexity of typical unitary generated by GUE Hamiltonian in the limit of large dimension.
To our knowledge, except from informal and not fully rigorous findings of \cite{ChaosRMT}, there are no results on complexity of GUE evolutions.

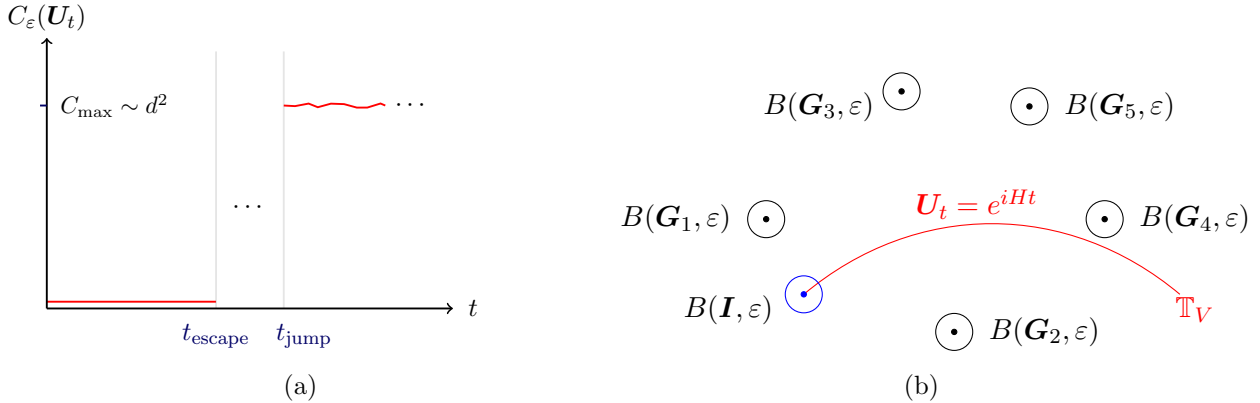
\begin{figure}
    \centering
\begin{subfigure}{0.5\textwidth}
    \definecolor{db}{rgb}{0,0,0.4}
    \scalebox{0.9}{
    \begin{tikzpicture}[scale=1]
    \node at (2.5,-0.4) {\textcolor{db}{$\tesc $}};
    \node at (3.8,-0.4) {\textcolor{db}{$\tlar $}};
    \draw[red,thick] (0,0.1) -- (2.5,0.1);
\draw[thick,color=gray,opacity=0.2] (2.5,0) -- (2.5,3.8);
    \node at (3,1.5) {$\cdots$};
    \draw[thick,color=gray,opacity=0.2]
    (3.5,0) -- (3.5,3.8);
    \draw[red,thick, decorate, decoration={random steps,segment length=5pt,amplitude=1pt}] (3.5,3) -- (5,3);
    \draw[thick,->] (0,0) -- (6,0);
    \draw[thick,->] (0,0) -- (0,4);
    \node at (6.3,0) {$t$};
    \node at (0,4.3) {$C_\varepsilon(\h{U}_t)$};
    \node at (5.4,3) {$\cdots$};
    \node at (1,3) {\small{$C_{\max} \sim d^2$}};
    \draw[thick,dashed,color=db] (-0.1,3) -- (0,3);
    \end{tikzpicture}
    }
    \caption{}
    \label{subfig:jump}
    \end{subfigure}%
    \begin{subfigure}{0.5\textwidth}
     \begin{tikzpicture}
\coordinate (O) at (0,0,0);
  \coordinate (A) at (5,0,0);
  \draw[color=red,-] (O) to [bend left=40] (A);
\node at (5.2,-0.2) {$\cred \mathbb{T}_V$};
\node at (2.3,1.2) {$\cred \U_t=e^{iHt}$};
\draw[blue] (0,0) circle (7pt);
\filldraw[blue] (0,0) circle (1pt);
\node at (-1,-0.2) {$B(\I,\varepsilon)$};
\draw (4,1) circle (7pt);
\filldraw (4,1) circle (1pt);
\node at (5.2,1) {$B(\h{G}_4,\varepsilon)$};
\draw (-0.5,1) circle (7pt);
\filldraw (-0.5,1) circle (1pt);
\node at (-1.7,1) {$B(\h{G}_1,\varepsilon)$};
\draw (2,-0.5) circle (7pt);
\filldraw (2,-0.5) circle (1pt);
\node at (3.2,-0.5) {$B(\h{G}_2,\varepsilon)$};
\draw (3,2.5) circle (7pt);
\filldraw (3,2.5) circle (1pt);
\node at (4.2,2.5) {$B(\h{G}_5,\varepsilon)$};
 \draw (1.3,2.7) circle (7pt);
\filldraw (1.3,2.7) circle (1pt);
\node at (0.2,2.5) {$B(\h{G}_3,\varepsilon)$};
\end{tikzpicture}
    \caption{}
    \label{subfig:proof}
    \end{subfigure}
    \caption{\small (a) Diagram presenting the 
    complexity jump, with threshold $t\sim \varepsilon$ appearing for evolutions under GUE Hamiltonian. Initially the complexity is zero and after a threshold it becomes nearly maximal. (b) Idea of our proof for unitary invariant ensembles proofs: we show that after the threshold time the evolution leaves the ball around identity, and that it avoids balls around low complexity unitaries $\h{G}_i\in\gwords{k}$. Technically we show that this avoidance holds with high probability for the whole torus $\mathbb{T}_V$ on which the evolution is localized ($V$ is Haar random unitary).
    }
    \label{fig:jump-proof}
\end{figure}

In this work we focus on two ensembles of random generators: GUE and the ensemble of diagonal Gaussian matrices conjugated by Haar random unitary transformations. Our main result is a striking complexity jump -- from trivial to nearly maximal. 
Namely, we show that in these models the time of achieving nearly maximal complexity is the same as the time of escaping the neighbourhood of identity -- see Fig.\ref{subfig:jump}. 
Since being in neighbourhood of identity means that the complexity of a unitary transformation is trivial, we have the aforementioned sudden jump.  Similar behavior is found for evolutions generated by Gaussian diagonal Hamiltonians and for states evolving according to Hamiltonians from the above unitary invariant evolutions.

To establish these results we rely heavily on structural properties of the above ensembles  and concentration
of measure techniques. This gives us a much finer control over the time evolution of
complexity compared to techniques previously employed in this context: high-degree
moments and frame potentials. 
The idea of our proof is depicted in Fig. \ref{subfig:proof}. We do this in two steps: we show that after threshold time $\tesc$ 
the evolution leaves the $\varepsilon$-ball around identity denoted as $B(\h{I},\varepsilon)$. On the other hand, we prove, that for all times, with  high probability the evolution avoids  $\varepsilon$-balls around low complexity
unitaries. The latter behaviour takes place for any unitarly invariant ensemble of Hamiltonians.
Technically, the crucial part of our approach is separating the treatment of the (random) eigenbasis and the (random) spectrum of the relevant Hamiltonian. The randomness of the eigenbasis can be handled using unitary invariance of the ensemble. Once the eigenbasis is chosen, the time evolution is confined to some specific torus embedded in the unitary group and is described by phases determined by the eigenvalues. These phases are, asymptotically as time goes to infinity, distributed uniformly and independently, so their expected asymptotic behavior is transparent. We stress that the resulting time evolution operator is asymptotically {\it not} distributed as a Haar random unitary since Haar random unitaries do not have independent eigenvalues. 

\subsection*{Significance of our results} 

Hamiltonians drawn from the GUE ensemble have exponentially many parameters and as such can only model strongly interacting systems. Nevertheless, the GUE ensemble has been successfully used in a diverse range of physical applications, usually to capture some form of chaoticity while remaining relatively tractable: quantum chaos \cite{appeal1}, thermodynamics \cite{appeal2}, many-body quantum physics (thermalization and equilibration \cite{appeal3}, objectivity \cite{appeal4}), High Energy Physics \cite{appeal5}, \cite{ChaosRMT}. The problem of rigorously analyzing late time behavior of a time independent Hamiltonian is notoriously difficult -- we believe that the GUE ensemble provides a good toy model and a necessary step for tackling the most physically relevant case (geometrically local Hamiltonians).

On the technical side, our work is to our knowledge the first one to prove complexity bounds without resorting to bounding high moments of the ensemble. Note that unlike e.g. random quantum circuits, an ensemble of unitary evolutions generated by time independent Hamiltonians will {\it not} form an exact $k$-design as $t \to \infty$ (as mentioned in the text, the limiting distribution is not the Haar measure), so there is no a priori reason for design-based methods to give meaningful bounds. Instead, we use more powerful concentration properties of the underlying ensemble and we believe this to be a fruitful future avenue for studying unitary evolutions generated by time-independent Hamiltonians (where, in any case, bounding moments rigorously is prohibitively difficult even in the GUE case).

\textbf{Structure of the paper:}  The paper is structured as follows. In Section \ref{sec:preliminaries} we introduce notation and basic mathematical setup used throughout the paper. Section \ref{sec:overview} contains outline of results, high level overview of  proof strategy and  discussion of open problems. The remaining sections are of technical nature. Section \ref{sec:unitary} deals with unitary complexity and builds up to the proof of main results for unitary complexity: Result \ref{result:GUE} and \ref{result:iid}. Subsections \ref{sec:haar} and \ref{sec:identity} deal with the case of unitary invariant ensembles, i.e. the $\GUE(d)$ and random basis i.i.d. Gaussian models. In the case of the diagonal i.i.d. Gaussian model (proof of Result \ref{result:diag}), a different approach is needed, which is presented in Subsection \ref{sec:fixed-basis}. Finally, Section \ref{sec:states} is concerned with evolution of state complexity and proves Result \ref{result:states}, with ideas similar to those used for unitary complexity. We have relegated proofs of some purely technical statements to the Appendix.

\textbf{Acknowledgements} We thank Adam Sawicki and Nick Hunter-Jones for helpful discussions.
MK and MO acknowledge financial support from the TEAM-NET project (contract no.\ POIR.04.04.00-00-17C1/18-00).
MH acknowledges support from the Foundation for Polish Science through an IRAP project co-financed by the EU within the Smart Growth Operational Programme (contract no.\ 2018/MAB/5).

\section{Preliminaries}\label{sec:preliminaries}

We will be concerned with complexity of objects defined on a finite dimensional Hilbert space $\H\simeq\C^d$ (for system of $n$ qubits we have $d=2^n$). The set of pure  quantum states $\S(d)$ consists of rank one projectors on normalised vectors in $\H$, denoted by $\psi:=\ketbra{\psi}{\psi}$. For every unitary operator $U$ on $\C^d$ we have a unitary channel $\h{U}$ defined by $\h{U}(\rho)=U \rho U^\dagger$. Let $\UU(d)$ denote the set of unitary channels in $\C^d$. Since, for a given unitary channel $\h{U}$ the global phase of its operator representative  $U$ is irrelevant, the set of unitary channels $\UU(d)$ is isomorphic to projective unitary group $P\unitary{d}=\unitary{d}/U(1)$. We will formulate our results in terms of standard distances on $\UU(d)$ and $\S(d)$: 
\begin{equation}\label{eq:distances}
    \ddiamond(\h{U}, \h{V}) :=  \norm{\h{U}-\h{V}}_{\diamond}\ , \ 
    \d(\psi, \phi) = \frac{1}{2}\norm{\psi-\phi}_{1} =\sqrt{ 1 - \tr(\psi \phi)}\ ,
\end{equation}
where $\norm{\cdot}_1$  and $\norm{\cdot}_\diamond$ denote trace and diamond norm respectively \cite{NielsenBook}. We will also use the following distances on $\UU(d)$ which are induced from norms on linear operators on $\C^d$:
\begin{equation}
    \d_\infty(\h{U}, \h{V}) := \inf_{\varphi \in [0,2\pi)}\norm{U - e^{i\varphi}V}_{\infty} \ ,\ \dhsproj(\h{U}, \h{V}) := \inf_{\varphi \in [0,2\pi)}\norm{U - e^{i\varphi}V}_{\HS}\ , 
\end{equation}
where $\norm{\cdot}_{\infty}$, $\norm{\cdot}_{\HS}$ denote the operator norm and the Hilbert-Schmidt norm, while $U,V$ are arbitrary unitary matrices that give rise to channels $\U$ and $\V$ (in what follows we would refer to such matrices as matrix representatives of unitary channels).  We have the following useful estimate for the diamond metric of unitary channels (\cite{OSH2020}, Proposition 6):
\begin{equation}\label{eq:diamond}
\d_\infty(\h{U}, \h{V})  \leq \ddiamond(\h{U}, \h{V}) \leq 2 \d_\infty(\h{U}, \h{V})
\end{equation}
In order to define state and unitary complexity we consider circuits formed from a universal discrete gate set $\gateset \subset \UU(d)$. In our context universality means only that the set of gate sequences  generated from $\gateset$ is dense in $\UU(d)$ (see however Remark \ref{rem:SKspecifics} for additional natural assumptions about the gate set, that make our results exhibit optimal scaling with  $d$).   Let $\gateset^r$ ($r=0,1,2,\ldots$) be the set of all size $r$ circuits built from $\gateset$. 

\begin{Definition}[Unitary complexity]\label{def:complexityUnitary}
For $\vep \in [0,1]$, we say that a unitary $\U$ has $\vep$-unitary complexity equal to $r$ if and only if $r=\min \left\{l: \exists \V\in\gateset^l ~\text{s.t.}~ \ddiamond(\U,\V)\leq \vep\right\}$, which we denote as $C_\vep(\U)=r$. 
\end{Definition}

\begin{Definition}[State complexity]\label{def:complexityState}
For $\vep \in [0,1]$, we say that a pure state $\psi$ has $\vep$-state complexity equal to $r$ if and only if $r=\min \left\{l: \exists \V\in\gateset^l ~\text{s.t.}~ \d(\psi,\V(\psi_0))\leq\vep \right\}$, which we denote as $C_\vep(\psi)=r$.
\end{Definition}

In the physical context we typically have $\H\simeq (\C^2)^{\otimes n}$ and  $\psi_0$ in the definition of state complexity is chosen to be an unentangled product state, for instance $\psi_0=\ketbra{0}^{\otimes n}$.

\textbf{Setting: }The main object of study of this paper are time evolutions governed by random Hamiltonians. We let the Hamiltonian $H$ to be drawn from some probability distribution on $d\times d$ Hermitian matrices and let $U_t = \exp(-itH)$ be the unitary time evolution operator associated to $H$. We will analyze in details the complexity of the corresponding unitary channel $\U_t$ and states $\psi_t:=\U_t (\psi_0)$ for the following models:

\begin{Definition}\label{def:gue}(Gaussian Unitary Ensemble ($\GUE$))
Fix $d$ and $\sigma > 0$. Let $X_{i, j}$ for $1\leq j \leq i \leq d$ and $Y_{i,j}$ for $j<i$ be independent Gaussians of mean $0$ and variance $\sigma^2$. We say that $H$ is drawn from $\GUE(d, \sigma^2)$ if $H$ is a hermitian matrix such that for every $i$ $ H_{ii} = X_{i,i}$ and $
H_{ij} = \frac{X_{i,j} + i Y_{i,j}}{\sqrt{2}}$ for $i>j$.
Note that $H \sim \GUE(d, \sigma^2)$ has the same distribution as $\sigma H', H' \sim \GUE(d, 1)$. We will write $\GUE(d)$ for $\GUE\left(d, \frac{1}{d}\right)$.
\end{Definition}

\begin{Definition}\label{def:iid}(Diagonal i.i.d Gaussian)
We say that $H$ is drawn from the diagonal i.i.d. Gaussian ensemble, denoted $\diaggauss{d}$, if $H$ is diagonal with diagonal elements $\lambda_i$ which are independent Gaussians of mean $0$ and variance $1$.
\end{Definition}

\begin{Definition}\label{def:iid-random-basis}(Random basis i.i.d Gaussian)
 We say that $H$ is drawn from the random basis i.i.d. Gaussian ensemble, denoted $\unitarygauss{d}$, if $H = UDU^{\dagger}$, where $U$ is Haar random unitary on $\C^d$ and $D$ is diagonal with elements $\lambda_i$ which are independent Gaussians of mean $0$ and variance $1$. 
\end{Definition}

\section{Overview of results}\label{sec:overview}

In this section we first outline our main findings about the complexity of random Hamiltonian evolution, then  give a high level overview our main proof techniques, and conclude with a list of open problems. Our typical result will be that for all times $t\leq\tesc$, $\U_t$ has with high probability trivial (zero) complexity, while after fixed slightly larger time $t>\tlar>\tesc$ it has nearly maximal complexity (scaling with $d^2$, see the discussion in Remark \ref{rem:SKspecifics}). 
 In the sharpest of the results, the ratio of the two timescales $\tlar/\tesc =\Theta(1)$, while individual timescales scale individually with accuracy parameter $\tesc\sim\vep$,  We thus obtain an extreme jump, which is illustrated in Fig. 
\ref{subfig:jump}.


In the results below
$U_t= e^{iHt}$, $\U_t$ is the corresponding unitary channel, and $\psi_t=\U_t(\psi_0)$, for a computationally trivial initial state $\psi_0$. The gateset $\gateset$ is fixed and $C,C',C''$ are absolute constants that may change from one result to another. On the other hand the parameter $\vep>0$ is either fixed (independent of $d$) or has to satisfy $\vep\gtrsim d^{-1/4}$ (a restriction needed for technical reasons) -- in particular, this allows the treatment of the physically relevant case for moderately small values of  $\vep=1/\mathrm{polylog}(d)=1/\mathrm{poly}(n)$ (with the latter equality valid for the system of $n$ qubits).


We start from:

\begin{result}[Complexity jump for GUE evolutions]\label{result:GUE}
Let $\tesc= C\vep$, $\tlar=C'\vep$ ($C'>C$). With high probability over $H\sim\GUE(d)$ 
\begin{enumerate}[(a)]
    \item \label{res1:trivial}
        For all times time $t\in[0, \tesc]$ complexity of evolution $\U_t$ is trivial: $
            C_{\vep}(\U_t)=0$ .
    \item \label{res1:maximal}
        For any fixed time $t> \tlar$  complexity satisfies
        \begin{equation}
             C_{\vep}(\U_t)\geq 
            \frac{C'' \vep^2 }{{\log\abs{\gateset}}} d^2
        \end{equation}  
which exhibits optimal scaling with $d$ due to Solovay-Kitaev compilation.
\end{enumerate}
\end{result}
\noindent The statement \eqref{res1:trivial} is proven in Proposition \ref{prop:identity} and follows from the property that for typical $H\sim\GUE(d)$  we have $\lambda_{max}-\lambda_{\min}=\Theta(1)$, while statement \eqref{res1:maximal} is proven in  Theorem \ref{th:complexity}.

\begin{remark}
For any $\vep' < \vep$ one of course has by definition $C_{\vep'}(U) \geq C_{\vep}(U)$ for any $U$. This observation can be used to reduce the undesired dependence on the above lower bound on $\vep$ as follows -- for any $\vep$ and $\vep' < \vep$ one has $ C_{\vep'}(\U_t)\gtrsim  \vep^2 d^2$ instead of $(\vep')^2 d^2$, for $ t > \tlar(\vep)$ (this is of course not guaranteed for $\tlar(\vep') < t <\tlar(\vep)$). In particular, if one is interested in complexity at some fixed constant time $t$, one can replace the dependence on $\vep^2$ in the bound by some constant.
\end{remark}

Analogous result holds in the case of random basis, i.i.d. Gaussian eigenvalues, yet with a less sudden jump. 
Namely, until $t\sim \vep/\sqrt{\log d}$ 
complexity is trivial, while after time $t\sim \vep$, it becomes near maximal. The ratio of these times exhibits a (sub) polynomial scaling with $n$ for the system of  $n$ qubits.

\begin{result}[Complexity jump for random basis  i.i.d. Gaussian Hamiltonians]\label{result:iid}
Let $\tesc= C\vep /\sqrt{\log(d)} $, $\tlar=C'\vep$. With high probability over the choie of random basis i.i.d. Gaussian Hamiltonian $H$ \begin{enumerate}[(a)]
        \item 
        \label{result2:trivial}
       For all times time $t\in[0, \tesc]$ the complexity of evolution $\U_t$ is trivial: $C_{\vep}(\U_t)=0$ .
        \item
        \label{result2:maximal}
       For any fixed time $t> \tlar$  the complexity satisfies 
       \begin{equation}
            C_{\vep}(\U_t)\geq 
            \frac{C'' \vep^2 }{{\log\abs{\gateset}}}d^2
       \end{equation}
        which exhibits optimal scalling with $d$ due to Solovay-Kitaev compilation.
    \end{enumerate}
\end{result}
\noindent   The part \eqref{result2:trivial} is proven in Proposition \ref{prop:identity} and follows from the property that typically $\lambda_{max}-\lambda_{\min} \sim \sqrt{\log(d)}$ for i.i.d. Gaussian Hamiltonians, while part \eqref{result2:maximal}
is proven in Theorem \ref{th:complexity}.  For discussion about the gap between $\tesc$ and $\tlar$, see Remark \ref{rm:gaussian-gap}.

Related results, albeit using different techniques and different scaling with respect to $\vep$, can be obtained for the random Gaussian Hamiltonian with fixed basis.

\begin{result}[Complexity bound for diagonal i.i.d. Gaussian Hamiltonians]
\label{result:diag}
Let $\tesc= C\vep /\sqrt{\log(d)} $, $\tlar=C'$.
With high probability over the choice of i.i.d. Gaussian Hamiltonian $H$ in the \emph{diagonal} basis
\begin{enumerate}[(a)]
    \item 
    \label{result3:trivial}
      For all times time $t\in[0, \tesc]$ complexity of evolution $\U_t$ is trivial: $C_{\vep}(\U_t)=0$ .
    \item 
    \label{result3:maximal}
      For any fixed time $t> \tlar$  complexity satisfies 
      \begin{equation}
          C_\vep(\U_t) \geq \frac{C''}{{\log\abs{\gateset}}}  d \log\left(\frac{1}{\vep}\right)\ .
      \end{equation}
        \end{enumerate}
\end{result}
\noindent The lower bound from Result  \ref{result:diag} \ref{result3:maximal} is near optimal in $d$ provided that eigenbasis of $H$ is the computational basis in $n$ qubit system.  This is due to the following upper bound valid for all unitaries diagonal in the computational basis: $ 
       C_{\vep}(\U)\leq  C d \log d \log^\gamma\left( \frac{1}{\vep} \right)$ (see Lemma \ref{lem:compBASIScompilation}). The statement \eqref{result3:trivial} is proven in Proposition \ref{prop:identity}, while \eqref{result3:maximal}, but without the $\log\left(\frac{1}{\vep}\right)$ scaling, is proved in Theorem \ref{th:complexity-fixed}. The proof of $\log\left(\frac{1}{\vep}\right)$ scaling requires a more particular technique and is given in Theorem \ref{th:complexity-fixed-eps}. Note that in the case of fixed basis model, in contrast to the GUE and random basis models, we only prove high complexity after \emph{constant} time (not $\sim \vep$), so we do not prove a sudden jump; see more discussion in Remark \ref{rm:fixed-gaussian}.

 Note that Result \ref{result:diag}\eqref{result3:trivial} and 
Result \ref{result:iid}\eqref{result2:trivial} are actually the same, since triviality of complexity means staying near the identity, which depends only on spectral properties and the two models have the same spectrum. 
Also, the above two results,
unlike Result \ref{result:GUE}, do not indicate whether after time $t\sim \frac{\vep}{\sqrt{\log d}}$ complexity becomes nonzero. Actually, this is the case -- namely, for any larger time, with high probability $U_t$ is outside an $\vep$-ball around the identity. This is proven in Proposition \ref{prop:gaussian}.

Similar picture as for unitary complexity can be drawn for state complexity.
\begin{result}[Complexity jumps for states evolving according to unitary invariant random Hamiltonians: GUE or random basis i.i.d. Gaussian]\label{result:states} Let $\tesc= C\vep$ for $H\sim\GUE(d)$ and  $\tesc= C\vep /\sqrt{\log(d)}$ for $H\sim\unitarygauss{d}$, $\tlar= C'\vep$.   Then, with high probability over the choice of $H$
    \begin{enumerate}[(a)]
        \item 
        \label{result4:trivial}
        For times $t\leq\tesc$ complexity of $\psi_t=\U_t(\psi_0)$ is trivial: $C_\vep(\psi_t)=0$.
        \item 
        \label{result4:maximal} For any fixed time $t> \tlar$  complexity satisfies
        \begin{equation}
            C_\vep(\psi_t) \geq \frac{C''\vep^2}{{\log\abs{\gateset}}}  d\ ,
        \end{equation}  
        which has near optimal scaling with $d$ due to the upper bound in Remark \ref{rem:SKspecifics}.
    \end{enumerate}
   
\end{result}
\noindent The statement \eqref{result4:trivial} follows from  application of Proposition \ref{prop:identity}, which states that for certain times the diamond distance of $\U_t$ from the identity is smaller than $\vep$. Due to the relation between diamond norm and operator norm \eqref{eq:diamond}, we obtain the same for states. The statement \eqref{result4:maximal} is proven in Theorem \ref{th:complexity-state}. In the case of state complexity, we do not consider the diagonal basis model, see Remark \ref{rm:no-fixed-basis}.

\begin{remark}
    Random Hamiltonians in considered models do not have a well prescribed notion of locality and hence their energy scale (and therefore also the time scale of unitary evolution) is not uniquely defined. In above definitions we adopted the conventions ensuring that with high probability $\norm{H}=\Theta(1)$ for GUE ensemble and $\norm{H}=\Theta(\sqrt{\log(d)})$ for the Gaussian ensemble. Importantly our results do not depend on these specific choices   of normalisation. What matters is the \emph{ratio} of the timescales: (i) $\tesc$ needed to escape the $\vep$ ball around identity channel $\I$ and (ii) $\tlar$ at which it is possible to claim essentially maximal complexity of $\U_t$. This ratio is not affected by the choice of normalisation for $H$.
\end{remark}

\begin{remark}\label{rem:SKspecifics}
    In order to claim optimality of our results with the dimension $d$ we need to have absolute upper bound for the complexity of unitaries and states. One way to realise this is to assume that we have $n$ qubit system, and our gateset consists of geometrically local universal gateset (acting nontrivially on a fixed number of qubits). In this case, assuming that the gateset is symmetric (i.e. containing every gate together with its inverse), the celebrated   Solovay-Kitaev algorithm \cite{Nielsen2005} gives the following absolute upper bound on complexity $    C_{\vep}(\U) \leq c(\gateset)d^2 \log\left( \frac{1}{\vep} \right)^\gamma$,
for some constant $c(\gateset)$ depending on the gateset and $\gamma\in[1,3)$. Furthermore, as shown in \cite{Harrow2002} we can chose $\gamma=1$ if our gate set enjoys the property of \emph{a spectral gap}   (which is conjectured \cite{Bourgain2011}to hold for arbitrary universal $\gateset$ and known to hold provided gates in $\G$ have algebraic entries). By similar techniques it is possible to obtain the following bound for state complexity $C_\vep (\psi) \leq c(\gateset) d \log\left (\frac{1}{\vep}\right)^\gamma$. When gateset $\gateset$ is not symmetric, currently known extensions \cite{BoulandOzols2017, BoulandTiron2021} of the Solovay-Kitaev algorithm give a worse exponent $\gamma$. Finally, stronger non-existential results \cite{Varju2013} on approximating sequences for general universal gatesets imply that for general non symmetric universal gatesets we can use above bounds for $\gamma\leq2$ (see also \cite{OSH2020} for detailed discussion). 
\end{remark}

\subsection*{Open problems} 

Below we highlight some of the open problems and future research directions related to findings of our paper.

First, it is natural to attempt to establish similar results for other classes of models of random  Hamiltonian evolutions. The most physically natural ones would be a random local Hamiltonians or models that exhibit structure or sparsity in some operator basis (such as the SYK model \cite{sykcomplexity} or the recently introduced random sparse Pauli models \cite{sparsePauli2023}).

Second, it would be interesting to study recurrence of time evolution, in a manner similar to \cite{OHH}. Note that the case of random Hamiltonians is more difficult than random quantum circuits, as the Hamiltonian is chosen once at the beginning, so the process $\{e^{iHt}\}_{t\geq 0}$ is non-Markovian. At the same time it seams plausible that by using techniques from \cite{Alhambra2022} one could get access to  \emph{average} recurrence times for unitary evolutions generated by Hamiltonians with \emph{generic} spectra. More generally, for fixed eigenbasis of $H$, time evolution $\exp(-itH)$ is equivalent to a dynamical system known as the linear flow on a torus. For this reason we expect that techniques of ergodic theory (for example quantitative results developed in \cite{Beck2017}) might be suitable to study recurrence properties of $\exp(-itH)$.

Finally, in our lower bounds on complexity, the parameter $\vep^2$ appears in the numerator. We believe it should be possible to remove it, although it may require more detailed knowledge about the distribution of unitaries generated from fundamental $\gateset$. It is also worth investigating whether purely geometric considerations (as opposed to using concentration of measure as we do in the paper) could give the optimal scaling with respect to $\vep$, namely $\log(1/\vep)$ -- see Appendix \ref{sec:appendix-Haar} for attempts in this direction.

\subsection*{Overview of proofs and techniques} 

We outline the high level strategy of the proofs and proof techniques.

\begin{remark}\label{rem:HScomplexity}
    For technical reasons it will be convenient for us to work with the approximate complexity $C^{\HS}_\vep (\U)$ defined in a manner analogous to that in Definition \ref{def:complexityUnitary} but using the Hilbert-Schmidt metric $\dhsproj$ as a distance. Note that the diameter of the set of unitaries with respect to the Hilbert-Schmidt metric scale as $\sqrt{d}$, so it is natural to consider complexity $C^{\HS}$ with parameter $\vep \sqrt{d}$, with $\vep$ independent of $d$.  
    
    By the inequality between the Hilbert-Schmidt and operator norm $ 
    \norm{A}_{\HS} \leq \sqrt{d} \norm{A}_{\infty}
    $
    it follows after taking infimum over phases and using Eq. \eqref{eq:diamond} that:
    for every $\U,\V$ \begin{equation}
        \dhsproj(\U, \V) \leq \sqrt{d} \cdot \ddiamond(\U, \V)\ .
    \end{equation}
    Hence, if for any $\h{G}\in\gateset^r$ we have $ \dhsproj(\h{U}, \h{G}) > \vep \sqrt{d}$, then also $\ddiamond(\h{U}, \h{G}) > \vep$. This proves that  $C_{\vep}(\U) \geq C^{\HS}_{\vep \sqrt{d}}(\U)$, so it suffices to provide lower bounds using the Hilbert-Schmidt metric.
\end{remark}

To bound complexity, we first use the union bound. For now let the distribution of $H$ be general. The probability below is with respect to the random choice of Hamiltonian $H$ and $U_t = \exp(itH)$
:\begin{equation}\label{eq:union-bound}
    \Pp_H(C^{HS}_{\vep \sqrt{d}}(\h{U}_t) \leq k ) = \Pp_H( \h{U}_t \in \bigcup_{\h{G} \in \gwords{k}} B(\h{G}, \vep\sqrt{d}) ) \leq \sum_{\h{G} \in \gwords{k}} \Pp_H\left( \h{U}_t \in  B(\h{G}, \vep\sqrt{d}) \right)
\end{equation}

We will now sketch the approach to estimating $\Pp\left( \U_t \in  B(\h{G}, \vep\sqrt{d}) \right)$ fleshed out in the following sections. Let us write $H = VDV^{\dagger}$, with $D$ real diagonal and $V$ unitary, where both $D$ and $V$ are in general random. We then have $U_t = V e^{iDt} V^{\dagger}$, where $e^{iDt}$ is diagonal with entries $(e^{\lambda_1 t}, \dots, e^{\lambda_d t})$, where $\lambda_i$ are the eigenvalues of $H$. This implies that once $V$ is chosen, the whole time evolution happens inside a $d$-dimensional torus $\mathbb{T}_{V}$ determined by $V$, where:
\begin{equation}\label{eq:torusV}
    \mathbb{T}_{V} = \{VAV^{\dagger} \ \vert \ A = \mathrm{Diag}(e^{i\varphi_1},\dots,e^{i\varphi_d}) , \varphi_i \in [0,2\pi)\}
\end{equation}
(note that the closure of the set $\{U_t\}_{t\geq 0}$ may have dimension strictly smaller than $d$ if the eigenvalues $\lambda_i$ are algebraically dependent)

Note that if the phases $(e^{\lambda_1 t}, \dots, e^{\lambda_d t})$ were distributed according to the Haar measure on the torus $T_V$, each probability appearing in the union bound would be of order $\sim \vep^{d}$, which would easily imply a complexity lower bound of order $\sim\frac{1}{\log\abs{\gateset}} d \log\left(\frac{1}{\vep}\right)$. Of course for general $H$ the phases will not be Haar random. Nevertheless, if the distribution of the eigenvalues of $H$ is continuous (as is the case in the ensembles we consider), it can be shown that the phase distribution will approach the Haar measure on the torus as $t\to\infty$ (see Appendix \ref{sec:appendix-Haar} for a more detailed discussion). Thus, we expect the same complexity lower bound of $\sim\frac{1}{\log\abs{\gateset}} d \log\left(\frac{1}{\vep}\right)$ to hold for large times. This explains heuristically why we expect the unitaries $\h{U}_t$ to have high complexity for large times.


We can obtain a much better $d^2$ scaling instead of just $d$ if the eigenbasis of $H$ is Haar random (for example, if $H$ is drawn from the $\GUE$ ensemble). If this is the case, one can show that for $\h{G}$ not too close to $\I$, the random torus $\mathbb{T}_V$ hosting the time evolution of $U_t$ will not intersect the ball $B(\h{G}, \vep)$ at all with very high probability. This will be made rigorous in Section \ref{sec:haar}. This part of the argument does not depend on the spectrum of $H$.

Then, the only term left in \eqref{eq:union-bound} will be $\h{G}=\I$. Note that since $\h{U}_0 = \h{I}$, by continuity for small times the channel $\U_t$ will typically be still close to the identity, so the bound on $\Pp\left( \U_t \in  B(\I, \vep) \right)$ will necessarily be time dependent and will depend on the spectrum of $H$ (but not on the eigenbasis). Intuitively, we expect it should take time $\sim \vep$ to leave the ball of radius $\vep$ around the identity with high probability. This will be made precise in Section \ref{sec:identity} for two models under considerations -- the GUE model and the random basis i.i.d. Gaussian model $\unitarygauss{d}$.

If the Haar randomness of the eigenbasis $V$ is not available (e.g. when $H$ always has a fixed eigenbasis or when the distribution of its eigenbasis is intractable), we are forced to deal with all the terms in \eqref{eq:union-bound}, not only $\h{G}=\I$. We analyze this problem for the diagonal i.i.d. Gaussian model $\diaggauss{d}$ in two different ways: in Section \ref{sec:fixed-basis} using probabilistic concentration inequalities and in Appendix \ref{sec:appendix-Haar} using purely geometric considerations. The former approach is more in line with other probabilistic techniques used in the paper, but gives worse scaling of the complexity lower bound with respect to $\vep$.

\begin{remark}
Our techniques can also be used to prove weaker complexity lower bounds in the case where the eigenbasis of $H$ is not Haar random, but comes from an approximate unitary $t$-design (and as such can be prepared e.g. by a relatively short random quantum circuit). Exponential concentration of measure on the unitary group needs then to be replaced by polynomial concentration coming from moment bounds.
\end{remark}

We end the overview by briefly commenting on the mathematical techniques used. The main tool for proving that certain desired properties hold with high probability is the concentration of measure \cite{meckes}. Roughly speaking, this means that if a quantity depends on a large number of random variables and not too strongly on any of them, it should be typically very close to its average value. The most familiar situation where such concentration occurs are sums of independent random variables (the law of large numbers setting), which are sharply peaked around their means. We employ this phenomenon in two ways. To handle the randomness of the eigenbasis of $H$, we use concentration of measure on the unitary group, which states that smooth functions of large unitary matrices are close to their expected values. Then, to analyze the time evolution on the torus we use concentration for functions of eigenvalues of a random Hamiltonian. In the case of the i.i.d. Gaussian ensembles this is simply concentration for independent variables, but for the GUE case we need technical statements coming from random matrix theory.  

\section{Unitary complexity}\label{sec:unitary}

\subsection{Avoiding 
 low complexity unitaries by random tori}\label{sec:haar}

Throughout this section we assume that $H = VDV^{\dagger}$ with $V$ distributed according to the Haar measure on $\unitary{d}$ and $D$ arbitrary diagonal. The time evolution $U_t$ happens inside $\mathbb{T}_V$ as defined in \eqref{eq:torusV}, which is a random torus since $V$ is Haar random. In this Section, we show that with high probability over the choice of $H$, this random torus will not intersect any of the balls around $\h{G}\in\gwords{k}$, apart from $\h{G}$ very close to the identity. This will allow us to focus only on the neighborhood of the identity, which is done in Section \ref{sec:identity}.

We first observe that since the time evolution $U_t$ happens inside $\mathbb{T}_V$, we have by conjugation invariance: 
\begin{equation}\label{eq:torus}
    \Pp\left( \h{U}_t \in  B(\h{G}, \vep \sqrt{d} ) \right) \leq \Pp_{V}\left( \mathbb{T}_V \cap B(\h{G}, \vep \sqrt{d}) \neq \emptyset \right) = \Pp_{V}\left( \mathbb{T} \cap B(\h{V}\h{G}\h{V}^{-1}, \vep \sqrt{d}) \neq \emptyset \right)  
\end{equation}
where $\mathbb{T}$ is the set of all diagonal unitary channels. Thus, to upper bound the right hand side of \eqref{eq:torus} it will suffice to show that with high probability over the choice of $V$, the distance from $VGV^{\dagger}$ to $\mathbb{T}$ is greater than $\vep \sqrt{d}$. 

To this end, let us define:
\begin{equation}\label{eq:dist}
    \dist(X, \mathbb{T}) := \inf_{D \in \mathbb{T}}\norm{X-D}_{\HS} = \inf_{D \in \mathbb{T}} \dhsproj(\h{X}, \h{D}) 
\end{equation}
where the last equality follows since $\mathbb{T}$ is closed under multiplication by $e^{i\varphi}$. We will first bound from below the expectation over $V$ of $\dist(VGV^{\dagger}, \mathbb{T})$ (Theorem \ref{th:expected-distance}) and then use concentration of measure to show that the distance is close to its expectation with high probability. 

\begin{lemma}\label{lm:udu}
For any unitary $X$ we have:
\begin{equation}
\dist(X, \mathbb{T})^2 = 2d - 2\sum_{i=1}^{d}\abs{X_{ii}}    
\end{equation}
\end{lemma}

\begin{proof}
It is straightforward to derive:
\begin{equation}
    \mathrm{dist}(X, \mathbb{T})^2 = 2d - 2 \sup_{D \in \mathbb{T}}\mathrm{Re}\tr(XD^{\dagger})
\end{equation}
We now claim that:
\begin{align}
\sup_{D \in \mathbb{T}} \mathrm{Re}\tr(XD^{\dagger}) = \sum_{i=1}^{d}\abs{X_{ii}}
\end{align}
Indeed, note that for any $D \in \mathbb{T}$:
\begin{align}
\mathrm{Re}\tr( X D^{\dagger}) \leq \abs{\tr(XD^{\dagger})} \leq
\sum_{i=1}^{d}\abs{(XD^{\dagger})_{ii}} = \sum_{i=1}^{d}\abs{(X)_{ii}}\abs{D^{\dagger}_{ii}} = \sum_{i=1}^{d}\abs{X_{ii}}
\end{align}
It is straightforward to see that the maximum value on the RHS can be obtained by putting $D_{ii} = \frac{X_{ii}}{\abs{X_{ii}}}$ if $X_{ii} \neq 0$ and $D_{ii}=1$ otherwise.
\end{proof}

A similar proof shows that for any unitary $X$ we have:
\begin{equation}\label{eq:dhs-id}
    \dhsproj(\h{X}, \h{I})^2 = 2d - 2\abs{\tr(X)}
\end{equation}

\begin{theorem}\label{th:expected-distance}
For any $\h{G} \in \gwords{k}$ we have:
\begin{equation}
    \E_{V}\dist(VGV^{\dagger}, \mathbb{T}) \geq \frac{1}{3} \cdot  \dhsproj(\h{G}, \h{I})
\end{equation}
(note that the projective Hilbert-Schmidt metric on the right hand side is necessary, since e.g. for $G=e^{i\varphi}I$ the left hand side is zero, but $\norm{G-I}_{\HS}>0$).
\end{theorem}

\begin{proof}
We will first bound the square of the expected distance. By Lemma \ref{lm:udu} we have:
\begin{equation}
    \E_V \dist(VGV^{\dagger}, \mathbb{T})^2 = 2d - 2\sum_{i=1}^{d}\E_{V} \abs{(VGV^{\dagger})_{ii}} = 2d - 2\sum_{i=1}^{d}\E_{V} \abs{\tr(\dyad{i}VGV^{\dagger})}
\end{equation}
To estimate the second term we use Jensen inequality:
\begin{equation}
\E_V \abs{\tr(\dyad{i}VGV^{\dagger})} \leq \left( \E_V \abs{\tr(\dyad{i}VGV^{\dagger})}^2\right)^{1/2} 
\end{equation}
and then perform the second moment computation which follows from Corollary 1 from \cite{quantum-avg-distance}:
\begin{equation}\label{eq:haar-average}
\E_V \abs{\tr(\dyad{i}VGV^{\dagger})}^2 = \frac{1}{d+1}\left(1 + \frac{\abs{\tr(G)}^2}{d}\right)
\end{equation}
Together this implies that:
\begin{equation}
    \E_V \mathrm{dist}(VGV^{\dagger}, \mathbb{T})^2 \geq 2d - 2d\sqrt{ \frac{1}{d+1}\left(1 + \frac{\abs{\tr(G)}^2}{d}\right) }
\end{equation}

By \eqref{eq:dhs-id}, we have $\abs{\tr(G)} = d - \frac{1}{2}\dhsproj(\h{G}, \h{I})^2$. Combining this with the inequality $\sqrt{1+x} \leq 1 + \frac{1}{2}x$ and $\dhsproj(\h{G}, \h{I})^4 \leq 2d \cdot \dhsproj(\h{G}, \h{I})^2$, after a short computation we end up with:
\begin{equation}\label{eq:square}
    \E_V \mathrm{dist}(VGV^{\dagger}, \mathbb{T})^2 \geq \frac{d}{2(d+1)} \cdot \dhsproj(\h{G}, \h{I})^2 \geq \frac{1}{3}\cdot \dhsproj(\h{G}, \h{I})^2
\end{equation}

Now, by \eqref{eq:dist} we have:
\begin{equation}
    \dist(VGV^{\dagger}, \mathbb{T}) = \inf_{D \in \mathbb{T}}\dhsproj(\h{V}\h{G}\h{V}^{-1}, \h{D}) \leq
    \dhsproj(\h{V}\h{G}\h{V}^{-1}, \h{I}) = \dhsproj(\h{G}, \h{I})
\end{equation}
which together with \eqref{eq:square} implies:
\begin{equation}
    \frac{1}{3}\cdot \dhsproj(\h{G}, \h{I})^2 \leq \E_{V}\dist(VGV^{\dagger}, \mathbb{T})^2 \leq \E_{V}\dist(VGV^{\dagger}, \mathbb{T}) \cdot  \dhsproj(\h{G}, \h{I})
\end{equation}
and the theorem is proved.
\end{proof}

Having bounded the expectation, we now move to the concentration of measure. To this end, we will need control over the Lipschitz constant:

\begin{lemma}\label{lm:lipschitz}
For any $G\in\unitary{d}$, the function $f: \unitary{d} \to \R$ defined as $f(V) = \mathrm{dist}(VGV^{\dagger},\mathbb{T})$ is Lipschitz with respect to the Hilbert-Schmidt norm with Lipschitz constant $2\norm{G-I}_{\infty}$.
\end{lemma}

\begin{proof}
We write $f = h \circ g$, where $g(V) = VGV^{\dagger}$ and $h(W) = \mathrm{dist}(W, \mathbb{T})$. First note that the function $\widetilde{g}(V) = VGV^{\dagger} - I = V(G-I)V^{\dagger}$ obviously has the same Lipschitz constant as $g$, so it suffices to bound the Lipschitz constant of $\widetilde{g}$. Using the properties of the Hilbert-Schmidt norm, we get:
\begin{align}
&\norm{\widetilde{g}(U) - \widetilde{g}(V)}_{\HS} = \norm{U(G-I)U^{\dagger} - V(G-I)V^{\dagger}}_{\HS} = \\
&\norm{U(G-I)(U^{\dagger}-V^{\dagger}) - (V-U)(G-I)V^{\dagger}}_{\HS} \leq \nonumber\\
&\norm{U(G-I)(U^{\dagger}-V^{\dagger})}_{\HS} +  \norm{ (V-U)(G-I)V^{\dagger}}_{\HS} \leq
2 \norm{G-I}_{\infty}\norm{U-V}_{\HS}
\end{align}
where in the last inequality we have used $\norm{UA}_{HS} = \norm{A}_{HS}$, valid for unitary $U$ and arbitrary $A$, and $\norm{AB}_{HS} \leq \norm{A}_{\infty} \norm{B}_{HS}$. This proves that $\widetilde{g}$, so also $g$, is Lipschitz with constant $2 \norm{G-I}_{\infty}$.

As for $h$, it is Lipschitz with constant $1$, which follows from the general fact that in any metric space $X$ and a set $A \subseteq X$ the function $x \mapsto \mathrm{dist}(x, A)$ is Lipschitz with constant $1$. Since $f=h\circ g$, it follows that $f$ is Lipschitz with constant $2\norm{G-I}_{\infty}$.
\end{proof}

We will use concentration of measure for $\unitary{d}$ in the following form (implied by \cite{meckes}, Theorem 5.17):
\begin{theorem}\label{th:concentration}
Let $F: \unitary{d} \to \R$ be Lipschitz with Lipschitz constant $L$ with respect to the Hilbert-Schmidt norm. Then for any $a > 0$:
\begin{align}\mathbb{P}_{U}\left( F(U) \geq \E_U F(U) + a \right) \leq \exp(-\frac{(d-2)a^2 }{ 12L^2}) 
\end{align}
\end{theorem}

We can now state and prove the main theorem of this section:

\begin{theorem}\label{th:main-sec1}
    Let $\h{G}$ be such that $\dhsproj(\h{G}, \h{I}) > 6 \vep \sqrt{d}$. Then for any $t \geq 0$ we have:
    \begin{equation}
    \Pp\left( \h{U_t} \in  B(\h{G}, \vep \sqrt{d}) \right) \leq \exp\left( - \frac{\vep^2 d^2}{384} \right)
    \end{equation}
\end{theorem}
\begin{proof}
By \eqref{eq:torus} it suffices to prove that:
\begin{equation}
    \Pp_{V}\left( \dist(VGV^{\dagger}, \mathbb{T}) \leq \vep \sqrt{d} \right) \leq \exp\left( - \frac{\vep^2 d^2}{384} \right)
\end{equation}
Let $F(V) = \mathrm{dist}(VGV^{\dagger}, \mathbb{T})$. By Theorem \ref{th:expected-distance}, we have:
\begin{equation}
    \E_{V} \mathrm{dist}(VGV^{\dagger}, \mathbb{T}) \geq \frac{1}{3} \dhsproj(\h{G}, \h{I}) \geq 2\cdot\vep \sqrt{d}
\end{equation}
By applying the concentration theorem (switching $F$ to $-F$) with $a=\vep\sqrt{d}$, invoking Lemma \ref{lm:lipschitz} and using $\norm{G-I}_{\infty} \leq 2$ we obtain:     
\begin{equation}
    \Pp_{V}\left( \dist(VGV^{\dagger}, \mathbb{T}) \leq \vep \sqrt{d} \right) \leq \exp\left( - \frac{\vep^2 (d-2)d}{48 \norm{G-I}^2_{\infty}} \right) \leq \exp\left( - \frac{\vep^2 d^2}{384} \right) 
\end{equation}
\end{proof}

\begin{remark}
    It would be desirable to improve the dependence of the probability bound on $\vep$, e.g. by removing the factor $\vep^2$. This could be possible by ensuring that $\norm{G - I}_{\infty}$, which controls the Lipschitz constant, is sufficiently small for all or most $\h{G} \in \gwords{k}$. However, available results about the distribution of unitaries from $\gwords{k}$ seem not to be strong enough to support such attempts.
\end{remark}

\subsection{Leaving the neighborhood of identity}\label{sec:identity}

In this Section, we analyze the probability that $\h{U}_t$ leaves the ball $B(\h{I}, \vep \sqrt{d})$ and prove that this happens with high probability as soon as $t \sim \vep$. The bulk of this Section is devoted to the GUE model. We then prove the same statement for the random basis i.i.d Gaussian model, which is considerably simpler. Arguments in this Section depend only on the spectrum of $H$ and not on its eigenbasis.

For now let the distribution of the Hamiltonian $H$ be general. By \eqref{eq:dhs-id}, we have:
\begin{equation}\label{eq:traceHL}
    \Pp( \dhsproj(\h{U}_t, \h{I}) < \vep \sqrt{d} ) =
    \Pp\left( \abs{\tr(U_t)} > (1-\frac{1}{2}\vep^2) d \right)
\end{equation}
where the probability is now only with respect to the distribution of eigenvalues of $U_t=e^{iHt}$. We now need to control the expected value and concentration  of $\tr (U_t)$. To this end, it is convenient to introduce the concept of spectral measure. Let $\lambda_1, \dots, \lambda_d$ be the eigenvalues of $H$ in arbitrary order. We define the spectral measure of $H$, denoted as $\mu_H$, to be the unique probability measure such that for any function $F$ of the eigenvalues we have:

\begin{equation}
    \int_{\R}F(\lambda)d\mu_H(\lambda) = \frac{1}{d}\E_{H}\sum_{i=1}^{d}F(\lambda_i)
\end{equation}
One can obtain $\mu_H$ by taking the average over the (random) empirical spectral measures $\mu = \frac{1}{d} \sum_{i=1}^{d}\delta_{\lambda_i}$.

We have:
\begin{equation}
    \frac{1}{d}\E_{H}\tr(U_t) = \frac{1}{d}\E_{H}\tr(e^{iHt}) =\frac{1}{d}\E_{H}\sum_{k=1}^{d}e^{i\lambda_k t} = \int_{\R}e^{i\lambda t}d\mu_H(\lambda) 
\end{equation}


We will now specify $H$ to be drawn from the GUE ensemble $\GUE(d)$ (c.f. Definition \ref{def:gue}). The scaling of variance as $\frac{1}{d}$ guarantees convergence of the distribution of eigenvalues to the Wigner semicircle law. For such $H$ we have the following Theorem, whose proof we defer to Appendix \ref{sec:appendixA}:
\begin{theorem}\label{th:wigner}
    Let $H_d \sim \GUE(d)$ and let $\mu_{sc}$ be the Wigner semicircle law with density $\rho_{sc}(x) = \frac{1}{2\pi}\sqrt{4-x^2} \id_{\{ \abs{x}\leq 2 \}}$. Then we have:
    \begin{equation}
        \lim_{d\to\infty}\int_{\R}e^{i\lambda t}d\mu_{H_{d}}(\lambda) = \int_{\R}e^{i\lambda t}d\mu_{sc}(\lambda) = \frac{J_1(2t)}{t}
    \end{equation}
    uniformly over $t$, where $J_1$ is the Bessel function of the first kind. The convergence rate is at least $O\left( \frac{\log d}{d} \right)$.
\end{theorem}
We note that the above statement is stronger than the classic convergence of GUE law to the Wigner semicircle law, since we require uniformity over all $t$.

\begin{theorem}\label{th:main-gue}
    Fix $\delta, \vep >0$. If $1/2 > \varepsilon > c d^{-1/4}$, for some constant $c$ depending on $\delta$, and $H \sim \GUE(d)$, then for any $t > C \vep$, where $C>0$ is an absolute constant, we have:
    \begin{equation}
        \Pp( \dhsproj(\h{U}_t, \h{I}) < \vep \sqrt{d} ) < \delta
    \end{equation}
\end{theorem}

\begin{proof}
    By \eqref{eq:traceHL}, we need to upper bound the probability:
    \begin{equation}
        \Pp\left( \abs{\tr(U_t)} > (1-\frac{1}{2}\vep^2) d \right)
    \end{equation}
    By standard properties of Bessel functions, we can pick an absolute constant $C$ such that $\abs{\frac{J_1(2t)}{t}} < 1 - 2 \vep^2$ if $t > C\vep$ (here we use $\vep < 1/2 < 1/\sqrt{2}$, so that $1-\frac{1}{2}\vep^2$ is bounded away from zero even if $\vep$ depends on $d$). Then, by Theorem \ref{th:wigner}, for sufficiently large $d$ we have $\abs{\int_{\R}e^{i\lambda t}d\mu_{H}(\lambda)} < 1 - \vep^2$ -- in fact, since the convergence rate is $\frac{\log d}{d}$, it suffices to have $\varepsilon > c d^{-1/4}$ for some $c$. This implies that $\abs{\E\tr(U_t)} < (1 - \vep^2)d$.

   We have by Chebyshev inequality:
    \begin{equation}\label{eq:chebyshev}
        \Pp\left( \abs{\tr(U_t)} > (1-\frac{1}{2}\vep^2) d \right) \leq
        \Pp\left( \abs{\tr(U_t) - \E\tr(U_t)} > \frac{1}{2}\vep^2 d \right) \leq
        \frac{4\Var\tr(U_t)}{\vep^4 d^2}
    \end{equation}
    where for a complex variable $Z$ the variance $\Var Z = \E\abs{Z}^2 - \abs{\E Z}^2$. The variance $\Var\tr(U_t)$ can be bounded for all $t$ using determinantal properties of GUE -- this is proved in Appendix \ref{sec:appendixA} in Theorem \ref{th:hermite}, which states that $\Var\tr(U_t)\leq d$. We thus have:
\begin{equation}
    \Pp\left( \abs{\tr(U_t)} > (1-\frac{1}{2}\vep^2) d \right) \leq 
    \frac{4}{\vep^4 d} < \delta
\end{equation}
which holds if $\vep > \left(\frac{4}{\delta}\right)^{1/4} d^{-1/4}$. This finishes the proof. We note that for small enough times $t$ the variance bound can be improved by writing:
\begin{equation}
    \Var\tr(U_t) = \Var\mathrm{Re}\tr(U_t) + \Var\mathrm{Im}\tr(U_t) = \Var\cos(Ht) + \Var\sin(Ht)
\end{equation}
and invoking the following variance bound (\cite{lytova}, Proposition 2.4; it is stated for GOE, but holds for $\GUE$ as well):

\begin{proposition}\label{prop:variance}
Let $H$ be a $d\times d$ $\GUE$ matrix with entries of variance $w^2$ and let $\varphi: \R \to \R$ be continuously differentiable with bounded derivative. Then:
\begin{align}
\Var\Tr \varphi(H) \leq 2 w^2 \sup_{\lambda \in \R}\abs{\varphi'(\lambda)}^2
\end{align}
\end{proposition}

In our case $w^2 = \frac{1}{d}$ and $\varphi(x)=\cos{xt}$ or $\sin(xt)$, which leads to $\Var\cos(Ht), \Var\sin(Ht) \leq \frac{2t^2}{d}$. 
\end{proof}

\begin{remark}
    The proof of Theorem \ref{th:main-gue} for the random basis i.i.d Gaussian model is considerably simpler, so we only sketch the necessary changes. It is straightforward to verify that if the eigenvalues $\lambda_i$ are independent standard Gaussians, then $\frac{1}{d}\E \tr(U_t) = e^{-t^2/2}$, which can be made $< 1 - 2 \vep^2$ if $t > C\vep$ for some $C>0$. One also readily computes that $\Var\tr(U_t) = d(1 - e^{-t^2})$, so the Chebyshev bound also holds. 
\end{remark}

\begin{remark}
The probability bounds used in the proof of Theorem \ref{th:main-gue} rely on estimating the second moment of $\tr(U_t)$ and are thus only inversely polynomial in $d$. For times $t = o(d)$ in the GUE case, or all times in the random basis i.i.d Gaussian case, exponential bounds are available using concentration of measure. We describe this approach in Appendix \ref{sec:appendixB}.
\end{remark}

We can now prove the main theorem describing the circuit complexity for $\GUE$ time evolution:

\begin{theorem}\label{th:complexity}
    Fix $\vep, \delta >0$ and the gateset $\gateset$. Let $H$ be drawn from the $\GUE(d)$ ensemble and assume $1/2 > \varepsilon > c d^{-1/4}$, for some constant $c$ depending on $\delta$. Then for any time $t > C\vep$, with $C$ an absolute constant, we have the following bound on the complexity of $U_t=e^{iHt}$ with respect to the Hilbert-Schmidt metric:
    \begin{equation}
    \Pp(  C^{\HS}_{\vep\sqrt{d}}(\h{U}_t) >  k ) > 1-\delta   
    \end{equation}
    with 
    \begin{equation}
    k = \frac{1}{{\log\abs{\gateset}}}\left(C' \vep^2 d^2 - \log(\frac{1}{\delta}) \right)    
    \end{equation}
    and $C'$ an absolute constant. The same result holds for random basis i.i.d Gaussian ensemble.
\end{theorem}

\begin{proof}
    We claim that for $t> C\vep$ and $d$ sufficiently large:
\begin{align}
\Pp\left( \h{U}_t \in \bigcup_{\h{G} \in \gwords{k}}B(\h{G}, \vep\sqrt{d}) \right) < \delta
\end{align}
Let us divide $\gwords{k} = \gwords{k}_{0} \cup \gwords{k}_{1}$, where $\gwords{k}_{0} := \{\h{G} \in \gwords{k} \ \vert \ \h{G} \in B(\h{I}, 6\vep\sqrt{d}) \}$ and $\gwords{k}_{1}$ contains the remaining elements. For every $\h{G} \in \gwords{k}_{0}$ we have $B(\h{G}, \vep\sqrt{d}) \subseteq B(\h{I}, 7\vep\sqrt{d})$, so:
\begin{align}
& \Pp\left( \h{U}_t \in \bigcup_{\h{G} \in \gwords{k}}B(\h{G}, \vep\sqrt{d}) \right) \leq \Pp\left( \h{U}_t \in B(\h{I}, 7\vep\sqrt{d}) \right) + \Pp\left( \h{U}_t \in \bigcup_{\h{G} \in \gwords{k}_{1}}B(\h{G}, \vep\sqrt{d}) \right) \leq \\
& \Pp\left( \h{U}_t \in B(\h{I}, 7\vep\sqrt{d})\right) + \sum_{\h{G} \in \gwords{k}_{1}}\Pp\left( \h{U}_t \in B(\h{G}, \vep\sqrt{d}) \right)
\end{align}

For every $\h{G} \in \mathcal{G}_{1}^{< k}$ we have $\dhsproj(\h{G},\h{I}) > 6\vep\sqrt{d}$, so we can invoke Theorem \ref{th:main-sec1}, to obtain:
\begin{equation}
    \sum_{\h{G} \in \gwords{k}_{1}}\Pp\left( \h{U}_t \in B(\h{G}, \vep\sqrt{d}) \right) \leq
    \abs{\gwords{k}} \exp\left( - \frac{\vep^2 d^2}{384} \right) < \frac{\delta}{2}
\end{equation}
if $k < \frac{C'}{\log\abs{\gateset}}\left(\vep^2 d^2 - \log(\frac{1}{\delta})\right)$ for appropriately chosen constant $C'$. Note that this part of the bound is time-independent.

It remains to bound $\Pp\left( \h{U}_t \in B(\h{I}, 7\vep\sqrt{d}) \right)$. To this end we simply invoke Theorem \ref{th:main-gue} and plug $\vep \to 7\vep, \delta \to 1/2\delta$.

The proof for the random basis i.i.d. Gaussian model proceeds in the same fashion.
\end{proof}

Theorem \ref{th:complexity} says that after a time $t > C \vep$, the operator $U_t$ typically has high complexity. We show a complementary result that $U_t$ stays in the neighborhood of the identity (so has trivial complexity equal to zero) for smaller times. In this sense, the complexity exhibits a sudden jump -- for sufficiently short time it has complexity zero and afterwards jumps to almost maximal complexity.

\begin{proposition}\label{prop:identity}
 Fix $\vep\leq \frac{\pi}{2}$. For times $t\leq \frac{\vep}{4}$ (GUE) or $t\leq \frac{\vep}{2\sqrt{2 \log(d)}}$ (i.i.d. Gaussian) the probability of leaving the $\vep$-ball around identity in the diamond norm for $\GUE(d)$ is bounded by $e^{-d^2}$ and for i.i.d. Gaussian by $\frac{1}{2d^2}$.
\end{proposition}
\begin{proof}
    From \cite{DiamondNormCharact} we know that the diamond distance between $\h{U}$ and $\h{I}$ is equal to the diameter of the set of eigenvalues of $U$. 
    Let $\phi_{min},\phi_{max}\in[-\pi,\pi)$ be ordered phases of eigenvalues of $U$, and assume that $\phi_{\max}-\phi_{\min} \leq \pi$. Then:
    \begin{align}
        \|\h{U} - \h{I}\|_\diamond=2 \sin\left(\frac{\phi_{\max}-\phi_{\min}}{2}\right) \leq  \phi_{\max}-\phi_{\min}
    \end{align}
    Let $\lambda_j$ be eigenvalues of $H$, with $\lambda_{\max},\lambda_{\min}$, being maximal and minimal eigenvalues respectively. Now, if  $t\|H\|_\infty\leq \pi/2$, 
    then we can write:
    \begin{align}
        \|\h{U} - \h{I}\|_\diamond \leq t (\lambda_{\max}-\lambda_{\min})\leq 2t \|H\|_\infty
    \end{align}
Then:
\begin{align}
    \Pp( \|\h{U} - \h{I}\|_\diamond \leq \vep)
    \geq\Pp( \|\h{U} - \h{I}\|_\diamond \leq \vep \ \cap \ t\|H\|_\infty\leq \pi/2)\geq \Pp (t\|H\|_\infty\leq \vep \ \cap \
    t\|H\|_\infty\leq \pi/2) 
\end{align}
so that for $\vep\leq \pi/2$  we get 
\begin{align}
    \Pp( \|\h{U} - \h{I}\|_\diamond \leq \vep) \geq \Pp (t\|H\|_\infty\leq \vep)
\end{align}
Using the following tail bound \cite{SzarekBook} for $\GUE(d)$:
\begin{align}
    \Pp(\|H\|_\infty\geq 2+a )\leq \frac12
    e^{-\frac12 a d^2}
\end{align} 
with any $a\geq 0$ 
and taking $t\leq \vep/4$
one gets 
\begin{align}
    \Pp(\|H\|_\infty\geq \frac{\vep}{t})
    =\Pp(\|H\|_\infty\geq 4)\leq \frac12e^{-d^2}
\end{align}
which ends the proof for GUE.
For the i.i.d. Gaussian model, we have the following tail bound:
\begin{align}
    \Pp(\|H\|_\infty\geq \sqrt{2 \log(d)}+a)\leq \frac12 e^{-\frac{a^2}{2}}.
\end{align}
Then taking $t \leq \frac{\vep}{2\sqrt{2 \log(d)}} $ we obtain the result.
\end{proof}

We end with a proposition concerning the random basis i.i.d. Gaussian model. The arguments used in Theorem \ref{th:complexity} work for times $t > C \vep$. However,  if we consider the diamond distance, in this model Proposition \ref{prop:identity} suggests that the threshold time to exit identity should behave like $\sim \frac{\vep}{\sqrt{\log d}}$. We show that this is in fact the case by working directly with the diamond metric (i.e. not using the Hilbert-Schmidt metric, as was done in previous arguments). The proof is presented in Appendix \ref{sec:appendixE}.

\begin{proposition}\label{prop:gaussian}
    Fix $\vep,\delta>0$ and let $\vep < 1/2$. Let $t > C\frac{\vep}{\sqrt{\log d}}$ for some constant $C>0$ depending on $\delta$. Then if $H$ is drawn from the random basis i.i.d. Gaussian model, we have:
    \begin{equation}
        \Pp\left(\ddiamond(\h{U}_t, \h{I}) > \vep\right) > 1 - \delta
    \end{equation}
\end{proposition}

\begin{remark}\label{rm:gaussian-gap}
The above Proposition concerns only leaving the ball around the identity and is not enough to show high complexity with respect to the diamond metric after time $t\sim \frac{\vep}{\sqrt{\log(d)}}$. The arguments from Section \ref{sec:haar}, used to eliminate all other balls, relied heavily on the use of Hilbert-Schmidt metric. In fact, the $\sim \exp(-\Theta(d^2))$ bound as in Theorem \ref{th:main-sec1} cannot hold in general in the diamond metric. Consider e.g. $G$ which is diagonal with eigenvalues $(-1, 1,1\dots,1)$. Then $UGU^{\dagger}=G$ for all unitaries $U$ which fix the first basis vector and the set of such unitaries has co-dimension $2d-1$ in $\unitary{d}$ -- this suggests we can only expect an $\sim \exp(-\Theta(d))$ type of bound if we want to use the same argument as in Theorem \ref{th:main-sec1} to exclude balls around $G \neq I$ with high probability.
\end{remark}

\subsection{Gaussian Hamiltonians in computational basis}\label{sec:fixed-basis}

\subsubsection{Lower bound on unitary complexity}

We now consider the Hamiltonian drawn from the diagonal i.i.d. Gaussian model, i.e. $H$ is diagonal with eigenvalues being i.i.d. standard Gaussians c.f. \ref{def:iid}. We will provide a lower bound on the complexity of $e^{iHt}$. In this case, the basis is fixed, so we cannot take advantage of the Haar randomness and concentration of measure on the unitary group to exclude all $\h{G} \in \gwords{k}$ but those close to the identity -- we need to deal with all of them. Note that the bounds used in Section \ref{sec:identity} rely on bounding the moments of $\abs{\tr(U_t)}$ and therefore cannot accommodate a union bound consisting of exponentially many in $d$ terms. To circumvent this problem we linearize $\abs{\tr(U_t)}$ to deal with its real and imaginary parts separately and then apply concentration for independent random variables. The bound on complexity will scale as $d$, not $d^2$ as in the random basis case, and in fact we show a nearly matching upper bound in Section \ref{sec:compilation}.

Let $\h{G} \in \gwords{k}$ be arbitrary. We aim to bound:
\begin{equation}\label{eq:trace}
    \Pp( \dhsproj(\h{U}_t, \h{G}) < \vep \sqrt{d} ) =
    \Pp\left( \abs{\tr(U_t G^{\dagger})} > (1-\frac{1}{2}\vep^2) d \right)
\end{equation}
Letting $\bra{k}G^{\dagger}\ket{k} = r_k e^{i\varphi_k}, \abs{r_k}\leq 1$ (the choice of global phase implicit in passing from $\h{U}_t, \h{G}$ to specific $U_t, G$ can be absorbed into $\varphi_k$) and using the fact that $U_t$ is diagonal with eigenvalues $e^{i\lambda_k t}$, we have:
\begin{equation}
    \abs{\tr(U_t G^{\dagger})}  = \abs{\sum_k r_k e^{i(\lambda_k t + \varphi_k)}}
\end{equation}
We make the following simple observation -- if $\abs{z} \geq c$, then at least one of: $\abs{\mathrm{Re}z} \geq \frac{1}{\sqrt{2}}c$ or $\abs{\mathrm{Im}z} \geq \frac{1}{\sqrt{2}}c$ must hold. By union bound, we thus obtain:
\begin{align}
&\Pp \left( \abs{\sum_k r_k e^{i(\lambda_k t + \varphi_k)}} > (1-\frac{1}{2}\vep^2) d     \right) \leq\\
&\Pp \left( \abs{ \mathrm{Re}(\sum_k r_k e^{i(\lambda_k t + \varphi_k)}) }> \frac{1}{\sqrt{2}}(1-\frac{1}{2}\vep^2) d     \right) +
\Pp \left(\abs{ \mathrm{Im}(\sum_k r_k e^{i(\lambda_k t + \varphi_k)})} > \frac{1}{\sqrt{2}}(1-\frac{1}{2}\vep^2) d     \right)
\end{align}

We focus on the real part first. We have:
\begin{align}
    &\Pp \left( \abs{ \mathrm{Re}(\sum_k r_k e^{i(\lambda_k t + \varphi_k)}) }> \frac{1}{\sqrt{2}}(1-\frac{1}{2}\vep^2) d     \right) \leq\\
    &\Pp \left(  \mathrm{Re}(\sum_k r_k e^{i(\lambda_k t + \varphi_k)})  > \frac{1}{\sqrt{2}}(1-\frac{1}{2}\vep^2) d     \right) + 
    \Pp \left(  \mathrm{Re}(\sum_k r_k e^{i(\lambda_k t + \varphi_k)}) < - \frac{1}{\sqrt{2}}(1-\frac{1}{2}\vep^2) d     \right)
\end{align}
We can write the first terms as:
\begin{equation}
    \Pp \left(  \sum_k r_k \cos(\lambda_k t + \varphi_k)  > \frac{1}{\sqrt{2}}(1-\frac{1}{2}\vep^2) d     \right)
.\end{equation}
Define random variables $X_k := r_k \cos(\lambda_k t + \varphi_k)$. Since the distribution of $\lambda_k$ is symmetric with respect to $0$, we have $\E\cos(\lambda_k t) = \E e^{i\lambda_k t} = e^{-t^2/2}$ and for the same reason $\E\sin(\lambda_k t) = 0$. Therefore, $\E X_k = r_k \cos\varphi_k e^{-t^2/2}$. It follows that $\E\sum_k X_k \leq d e^{-t^2/2}$. By choosing $t$ large enough we can make $e^{-t^2/2} < \frac{1}{\sqrt{2}}(1-\frac{1}{2}\vep^2) - \frac{1}{2\sqrt{2}}$, therefore:
\begin{equation}
    \Pp \left(  \sum_k X_k  > \frac{1}{\sqrt{2}}(1-\frac{1}{2}\vep^2) d     \right) \leq
        \Pp \left(  \sum_k X_k  -\E\sum_k X_k > \frac{1}{2\sqrt{2}}d     \right) 
\end{equation}
We are now in position to use concentration inequality for bounded independent variables in the form of Hoeffding bound:
\begin{proposition}[Hoeffding's inequality]
Let $X_i$ be independent random variables such that almost surely $a_i \leq X_i \leq b_i$. Let $S_n = X_1 + \dots + X_n$. Then for all $s>0$:
\begin{align}
\Pp\left( S_n - \E S_n \geq s \right) \leq \exp\left( -\frac{s^2}{\sum_{i=1}^{n}(b_i-a_i)^2} \right)
\end{align}
\end{proposition}
In our case $a_i=1, b_i=-1$, so the final inequality reads:
\begin{equation}
 \Pp \left(  \sum_k X_k  -\E\sum_k X_k > \frac{1}{2\sqrt{2}}d     \right)    \leq \exp\left(-\frac{d}{32}\right)
\end{equation}
Bounding the other term with the real part and the terms with imaginary part proceeds in the same way. By the same reasoning as in the proof of Theorem \ref{th:complexity}, we have arrived at the following complexity bound. Note that there is no dependence of the complexity on $\vep^2$, but at the cost of considering times $t$ greater than a constant instead of $\sim \vep$.

\begin{theorem}\label{th:complexity-fixed}
        Fix $\vep, \delta >0$ and the gateset $\gateset$. If $H$ is drawn from the diagonal i.i.d. Gaussian model, then for any time $t > C$, with $C$ an absolute constant, we have:
    \begin{equation}
    \Pp(  C^{HS}_{\vep\sqrt{d}}(\h{U}_t) >  k ) > 1-\delta   
    \end{equation}
    with 
    \begin{equation}
    k = \frac{1}{{\log\abs{\gateset}}} \left(C' d - \log(\frac{1}{\delta}) \right)    
    \end{equation}
    and $C'$ an absolute constant. 
\end{theorem}

 \begin{remark}\label{rm:fixed-gaussian}
     The above Theorem shows high complexity only after constant time. This is in contrast with unitary invariant ensembles, where high complexity occurs right after leaving the ball around identity, i.e. at time $t\sim\vep$. In the absence of random basis, we are forced to consider all balls in the union bound. If we do not know anything about their distribution, we have to handle their possible worst case placement, e.g. all balls clustered around identity so that it takes constant time to even leave that cluster with high probability. Excluding such cases and proving that time $t\sim\vep$ is sufficient would require a rather fine grained control of the distribution of unitaries in $\gwords{k}$ (= the distribution of a random walk on the unitary group with steps from $\gateset$). 
 \end{remark}

\subsubsection{Upper bound on complexity via explicit compilation for qubits}\label{sec:compilation}

For the upper bound, we specialize the Hilbert space to the case of an $n$ qubit system, so that $d = 2^n$. We first note that any Hamiltonian acting on $n$ qubits diagonal in the computational basis can be written as:
\begin{align}
    H=\sum_\alpha  \lambda_\alpha Z^\alpha
\end{align}
where the sum runs over $n$-tuples $\alpha=(\alpha_1,\ldots, \alpha_n)$, $\alpha_j=0,1$  and 
\begin{align}
Z^\alpha=Z_1^{\alpha_1}\ldots Z_n^{\alpha_n}
\end{align} 
with $Z_j$ being Pauli matrix on $j$'th qubit. 

\begin{lemma}\label{lem:compBASIScompilation}
Consider an arbitrary Hamiltonian $H$ acting on $n$  qubits which is  diagonal in computation basis. Let the gateset contain $CNOT's$ and $Z$ gates. Then the $\vep$-complexity of the unitary $U_t=e^{i t H }$ (defined with respect to the diamond metric) satisfies:
    \begin{align}
       C_{\vep}(U_t)\leq C n 2^n \log^c\left( \frac{1}{\vep} \right) = C d \log d \log^c\left( \frac{1}{\vep} \right)
    \end{align}
    where $c$ and $C$ are absolute constants.
\end{lemma}
\begin{proof}
We first claim that:
\begin{align}
e^{it Z^\alpha }=V_\alpha e^{it Z_1 } V_\alpha^\dagger
\end{align}
where $V_\alpha$ is composed out of at most $4n$ gates. Since for any unitary we have $Ue^X U^\dagger= e^{UXU^\dagger}$, it is enough to find $V_\alpha$ such that 
    \begin{align}
        Z^\alpha=V_\alpha Z_1 V_\alpha^\dagger.
    \end{align}
    We construct $V_\alpha$ as follows. 
    Since CNOT's propagate $Z$-errors backwards,
    we sandwich $Z_1$ with a cascade of CNOT's to produce $Z_1 \ldots Z_n$. Then we need to remove $Z$'s in places where $\alpha_j=0$.
    We do this by adding another CNOT's with source qubits coinciding with the places where we do not want Z's. Such $V_\alpha$ requires no more than $2n$ CNOTs. The circuit is illustrated in Fig. \ref{fig:cnots}.
    \begin{figure}
        \centering
        \includegraphics[width=14cm]{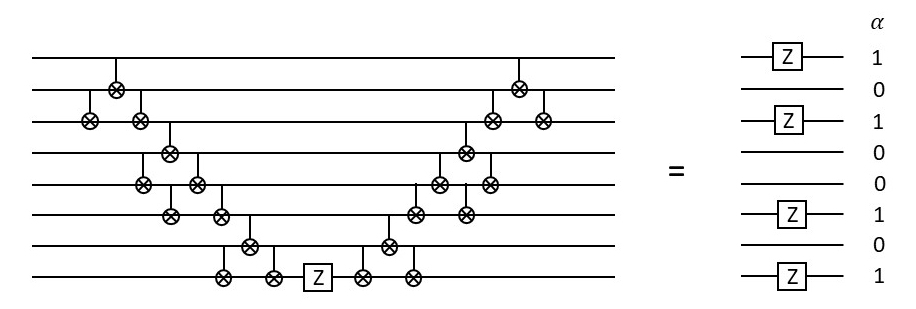}
        \caption{Compilation of $Z^\alpha$ from $Z_1$ and c-not gates.}
        \label{fig:cnots}
    \end{figure}
    
Now, to compile $e^{itH}$ we proceed as follows. 
First we compile $e^{itZ_1}$ up to precision $\delta$ using $k=\ln^c\frac1\delta$ gates, obtaining $U_0$
satisfying $\|U_0-e^{itZ_1}\|_{\infty}\leq \delta$ (as follows from the Solovay-Kitaev theorem).
Next we sandwich $U_0$ with $V_\alpha,V_\alpha^\dagger$,
obtaining  $U_\alpha=V_\alpha U_0V_\alpha^\dagger$
which is $\delta$ approximation of $e^{it Z^\alpha}$ and costs $4n+k$  gates. 
Finally, since $e^{itH}=\Pi_\alpha e^{it Z^\alpha}$, 
we compose $U_\alpha$'s, 
which by a telescopic estimate, satisfies
\begin{align}
    \|\Pi_\alpha e^{it Z_\alpha} - U_{2^n-1}\ldots U_0\|_{\infty}\leq 2^n \delta
\end{align}
where we have written $\alpha$'s in a unary way. 
It remains to choose $k$ such that  the above norm is smaller than $\vep$. This is done by $k=n  \log^c2 +\log^c\frac1\vep$.
Overall, the number of needed gates is 
$K=2^n \times 4n\times k=4n2^n(n  \log^c2 +\log^c\frac1\vep)$,
or it is enough to take $K= C n 2^n  \log^c\frac1\vep$,
 where $C$ is absolute constant. We have obtained an approximation in the operator norm, which implies approximation in the diamond metric. Hence: 
\begin{align}
    C_\vep(e^{it H}) \leq C n 2^n  \log^c\frac1\vep.\end{align}
\end{proof}

\section{State complexity}\label{sec:states}

The general strategy for lower bounding $C_{\vep}(\U_t(\psi_0))$ will be the same as in the unitary case -- we first aim to exploit the Haar randomness of the eigenbasis of $H$ and then upper bound the probability of staying close to the initial state $\psi_0$. The reader is advised to first read Section \ref{sec:unitary}.

\begin{remark}\label{rm:no-fixed-basis}
    In the case of state complexity, we do not consider the diagonal Gaussian model $\diaggauss{d}$ as the results would depend very strongly on the choice of the initial state. For example, if the state $\psi_0$ is chosen to be one of the computational basis states $\ket{k}$, which are the eigenstates of the Hamiltonian $H$, it will not evolve at all, so the complexity of $\U_t(\psi_0)$ will always be trivial. In general, if we write:
    \begin{equation}
        \ket{\psi_0} = \sum_{k}c_k \ket{k}
    \end{equation}
    then, for example, analyzing the trace distance between $\psi_0$ and $\U_t(\psi_0)$ is equivalent to analyzing:
    \begin{equation}\label{eq:weighted-trace}
        \bra{\psi_0}U_t\ket{\psi_0} = \sum_k \abs{c_k}^2 e^{i\lambda_k t}
    \end{equation}
    If $\abs{c_k}^2 = \frac{1}{d}$ for all $k$, we are reduced to analyzing the trace of $U_t$, which was done in previous sections. For general coefficients $c_k$ the probabilistic properties of \eqref{eq:weighted-trace} will depend on some measure of uniformity of $c_k$.
\end{remark}

\subsection{Random tori in state spaces}
As in the unitary case, we first aim to exploit the Haar randomness of the basis to eliminate from consideration all balls but the one at the initial state $\ket{0}$. For a fixed state $\ket{\phi}$, let us define:

\begin{equation}\label{eq:torus1}
    T_{\ket{\phi}} := \left\{ \ket{\alpha} \ \vert  \ \forall k=1,\dots,d \ \abs{\braket{k \vert \alpha}} = \abs{\braket{k \vert \phi}}  \right\}
\end{equation}

Let $H=V^{\dagger}DV$, with $V$ a Haar random unitary. For a fixed $\ket{\phi} \in \states{k}$ we have:
\begin{equation}
    \d(\ket{\phi},U_t\ket{0})^2 = 1 - \abs{\bra{\phi}V^{\dagger} e^{iDt} V\ket{0}}^2
\end{equation}
Note that:
\begin{equation}
e^{iDt} V\ket{0} = \sum_{k}e^{i\lambda_k t} \dyad{k}{k}V\ket{0}
\end{equation}
Thus, the states of the form $e^{iDt} V\ket{0}$ lie inside a (random) $d$-dimensional torus $T_{V\ket{0}}$ as defined in \eqref{eq:torus1}. We will prove that for any state $\ket{\phi}$ which is at least $\vep$ away from $\ket{0}$, the probability over $V$ that $V\ket{\phi}$ is $\vep$-close to $T_{V\ket{0}}$ is exponentially small (note that the same $V$ occurs twice in this statement), and thus $U_t\ket{0}$ avoids the the $\vep$-ball around $\ket{0}$. This allows us to focus only on the neighborhood of $\ket{0}$. The logic of the argument is essentially the same as in the operator case.

Define:
\begin{equation}
    \dist(\ket{\psi}, T_{\ket{\phi}}) := \inf_{\ket{\alpha} \in T_{\ket{\phi}}} \d(\ket{\psi}, \ket{\alpha})
\end{equation}

\begin{lemma}\label{lm:distance-from-torus}
Fix two states $\ket{\psi},\ket{\phi}$. We then have:
\begin{equation}
    \dist(\ket{\psi},T_{\ket{\phi}})^2 = 1 - \left( \sum_k \abs{\braket{k\vert \psi}}\cdot\abs{\braket{k\vert \phi}}\right)^2 
\end{equation}
\end{lemma}

\begin{proof}
    By definition, for any $\ket{\alpha} \in T_{\ket{\phi}}$ we have $\ket{\alpha} = \sum_{k} e^{i\varphi_k}\ket{k}\braket{k\vert \phi}$, so:
    \begin{equation}
        \dist(\ket{\psi},T_{\ket{\phi}})^2 = 1 - \sup_{\ket{\alpha} \in T_{\ket{\phi} }} \abs{\braket{\psi\vert \alpha}}^2 = 
        1 - \sup_{\varphi_k} \abs{\sum_k  e^{i\varphi_k}\braket{\phi\vert k}\braket{k \vert \psi}  }^2
    \end{equation}
    and clearly the $\sup$ is attained for the choice $\varphi_k$ such that $e^{i\varphi_k}\braket{\phi \vert k}\braket{k \vert \psi} = \abs{\braket{\phi \vert k}\braket{k \vert \psi}}$, which finishes the proof.
\end{proof}

As in the unitary case, we would first like to estimate $\E_{V}\dist(V\ket{0}, T_{V\ket{\phi}})$ in terms of $d(\ket{0}, \ket{\phi})$. Curiously, trying to estimate:
\begin{equation}
    \E_V \dist(V\ket{0}, T_{V\ket{\phi}})^2 = 1 - \E_V \left( \sum_k \abs{\bra{k}V\ket{0}}\cdot \abs{\bra{k}V\ket{\phi}} \right)^2 
\end{equation}
by reducing the expression to polynomials in $\abs{\bra{k}V\ket{0}}^2, \abs{\bra{k}V\ket{\phi}}^2$ (as was done in the unitary case) does not result in useful bounds. We are therefore forced to analyze the expressions $\E_V \abs{\bra{k}V\ket{0}}\cdot \abs{\bra{k}V\ket{\phi}}$ directly, which is more difficult since they are not polynomials. To achieve this, note that $V\ket{k}$ is a Haar random state from the complex sphere. It is well known that coordinates of such vectors are well approximated by complex Gaussian variables. This is formalized in the following Lemma. Note that by the Haar randomness of $V$ we can without loss of generality only consider states of the form $\ket{\phi} = \alpha\ket{0} + \beta\ket{1}$.

\begin{lemma}\label{lm:gaussian-approximation}
    Let $\ket{\phi} = \alpha\ket{0} + \beta\ket{1}$, with $\abs{\alpha}^2+\abs{\beta}^2=1$. Let $V$ be a $d \times d$ Haar random unitary and let $X,Y$ be independent standard complex Gaussians, i.e. $X, Y \sim a + ib$, where $a,b$ are real Gaussians of mean zero and variance $1/2$. Then we have:
    \begin{equation}
        \lim_{d\to\infty} d\cdot\E_{V}\abs{\bra{k} V \ket{0}}\abs{\bra{k} V \ket{\phi}} = \E\abs{X}\abs{\alpha  X +  \beta  Y   }
    \end{equation}
    uniformly over the choice of $\alpha, \beta$ and with convergence rate at least $d^{-1/2}$.
\end{lemma} 

\begin{proof}
    Note that for any $\ket{k}$ the state $V\ket{k}$ is a Haar random state. It is known \cite{diaconis}that if $X$ is a randomly distributed point on the real $n$-dimensional sphere of radius $\sqrt{n}$ and $Z$ is a standard real Gaussian random vector, then for any $k \leq n$:
    \begin{equation}\label{eq:diaconis}
        d_{TV}((X_1, \dots, X_k), (Z_1, \dots, Z_k)) \leq \frac{2(k+3)}{n-k-3}
    \end{equation}
    where $d_{TV}$ is the total variation distance. The bound immediately applies also to the complex case. Applying this bound for $k=2$, we obtain that the joint law of $\sqrt{d}(\bra{k} V \ket{0}, \bra{k} V \ket{1})$ converges in total variation distance to $(X, Y)$, where $X, Y$ are standard independent complex Gaussians. 

    By unitary invariance, without loss of generality we can write $\ket{\phi} = \alpha\ket{0} + \beta\ket{1}$, where $\abs{\alpha}^2 + \abs{\beta}^2 = 1$. Let $\mu_d$ denote the joint law of $\sqrt{d}(\bra{k} V \ket{0}, \bra{k} V \ket{1})$, let $\mu$ denote the joint law of $(X, Y)$ and let $f_{\alpha}(x,y)=\abs{\alpha \abs{x}^2 + \beta xy}$. We are left with proving uniformly over $\alpha$ that:
    \begin{equation}
        \lim_{d\to\infty}\E_{\mu_d}f_{\alpha} = \E_{\mu}f_{\alpha}
    \end{equation}
    The argument below is standard, but since we need to ensure uniformity over $\alpha$ and the functions $f_{\alpha}$ are not bounded, we spell it out in detail. Pick $K>0$ and write:
    \begin{align}\label{eq:K}
    &\abs{\E_{\mu_d}f_{\alpha} - \E_{\mu}f_{\alpha}} \leq \\
    &\abs{\E_{\mu_d}f_{\alpha} - \E_{\mu_d}f_{\alpha}\id_{\{ f_{\alpha} \leq K\}}} +
    \abs{\E_{\mu_d}f_{\alpha}\id_{\{ f_{\alpha} \leq K\}} - \E_{\mu}f_{\alpha}\id_{\{ f_{\alpha} \leq K\}}} +
    \abs{\E_{\mu}f_{\alpha}\id_{\{ f_{\alpha} \leq K\}} - \E_{\mu}f_{\alpha}} = \\
    & \abs{ \E_{\mu_d}f_{\alpha}\id_{\{ f_{\alpha} \geq K\}}} +
    \abs{\E_{\mu_d}f_{\alpha}\id_{\{ f_{\alpha} \leq K\}} - \E_{\mu}f_{\alpha}\id_{\{ f_{\alpha} \leq K\}}} +
    \abs{\E_{\mu}f_{\alpha}\id_{\{ f_{\alpha} \geq K\}} } 
    \end{align}
    We first bound the second term. From the dual formulation of the total variation distance, we know that:
    \begin{equation}
        d_{TV}(\mu_d, \mu) = \frac{1}{2}\sup_{g}\abs{\E_{\mu_d}g - \E_{\mu}g}
    \end{equation}
    where the supremum is taken over all bounded measurable function $g$ such that $\abs{g} \leq 1$. This easily implies that:
    \begin{equation}\label{eq:K-TV}
        \abs{\E_{\mu_d}f_{\alpha}\id_{\{ f_{\alpha} \leq K\}} - \E_{\mu}f_{\alpha}\id_{\{ f_{\alpha} \leq K\}}} \leq K \cdot d_{TV}(\mu_d, \mu)
    \end{equation}
    To bound the first and third terms in \eqref{eq:K}, we use the bound:
    \begin{equation}\label{eq:K-bound}
        \E_{\mu_d}f_{\alpha}\id_{\{ f_{\alpha} \geq K\}} \leq \frac{\E_{\mu_d} f^2 }{K}
    \end{equation}
    Note that $\E_{\mu_d} f^2 \leq 1$. Indeed, we have by Cauchy-Schwartz:
    \begin{equation}
        \E_{\mu_d} f^2 = d\cdot\E_{V}\abs{\bra{k} V \ket{0}}\abs{\bra{k} V \ket{\psi}} \leq
        d (\E_{V}\abs{\bra{k} V \ket{0}}^2)^{1/2}(\E_{V}\abs{\bra{k} V \ket{\psi}}^2)^{1/2} = 1
    \end{equation}
    A similar bound holds for $\mu$. Inserting bounds \eqref{eq:K-TV} and \eqref{eq:K-bound} into \eqref{eq:K}, we can choose $K$ large enough and then $d$ large enough so that $\abs{\E_{\mu_d}f_{\alpha} - \E_{\mu}f_{\alpha}}$ is arbitrarily small, which finishes the proof. Since the convergence rate in \eqref{eq:diaconis} is explicit, one easily checks that for a given error $\vep$ it suffices to take $K \sim \frac{1}{\vep}$ and then $d \sim \vep^{-2}$, which overall implies convergence rate $d^{-1/2}$.
\end{proof}

We need to estimate the Gaussian average $\E\abs{X}\abs{\alpha  X +  \beta  Y   }$. Such an estimate is supplied by the following Lemma, whose proof we defer to Appendix \ref{sec:appendixD}:

\begin{lemma}\label{lm:gaussian-average}
    Let $X, Y$ be independent standard complex Gaussians. Let $\abs{\alpha}^2 + \abs{\beta}^2 = 1$. The following inequalities hold for some constant $c>0, \beta_0 \in (0,1)$:
    \begin{equation}
       1 - (\E\abs{X}\abs{\alpha X +  \beta  Y   })^2 \geq c \min\{\abs{\beta}^2 , \beta_0^2 \}
    \end{equation}
\end{lemma}

\begin{lemma}\label{lm:f-lipschitz}
The function $f: \unitary{d}\to\R$, defined as:
\begin{equation}
    f(U) = \sum_k\abs{\bra{k} U \ket{0}}\abs{\bra{k} U \ket{\phi}}
\end{equation}
is Lipschitz with respect to the Hilbert-Schmidt norm with Lipschitz constant $1$.
\end{lemma}

\begin{proof}
    By Lemma \ref{lm:distance-from-torus} we have:
    \begin{equation}
        f(U) = \sup_{\ket{\alpha} \in T_{U\ket{\phi}}} \abs{ \bra{\alpha} U \ket{0} }
    \end{equation}
    For any fixed $\ket{\alpha} \in T_{U\ket{\phi}}$, the function $f_{\ket{\alpha}}(U) := \abs{ \bra{\alpha} U \ket{0} }$ is Lipschitz with constant $1$. Since $f(U) = \sup_{\ket{\alpha}}f_{\ket{\alpha}}(U)$, it is also Lipschitz with constant $1$ as a supremum of uniformly Lipschitz functions.
\end{proof}

\begin{theorem}\label{th:main-state}
    Let $c, \beta_0$ be as in Lemma \ref{lm:gaussian-average} and let $C_0 = \frac{2}{\sqrt{c}}$. Let $\ket{\phi}$ be such that $d(\ket{\phi},\ket{0}) > C_0\vep $ and let $\vep < \frac{\beta_0}{C_0}$. Then for some absolute constant $c'$ and $\vep > c' d^{-1/4}$ we have that for any $t \geq 0$:
    \begin{equation}
    \Pp\left( U_t\ket{0} \in  B(\ket{\phi}, \vep) \right) \leq \exp\left( - \frac{\vep^2 d}{6} \right)
    \end{equation}
\end{theorem}

\begin{proof}
    To prove the theorem it suffices to upper bound $\Pp(\dist(V\ket{0}, T_{V\ket{\phi}}) < \vep  )$. Without loss of generality assume that $\ket{\phi} = \alpha\ket{0} + \beta\ket{1}$, where $\abs{\alpha}^2 + \abs{\beta}^2=1$. Note that $\abs{\beta} = d(\ket{\phi},\ket{0}) > C_0 \vep  $. Let 
    \begin{equation}
    Z_d = \sum_k\abs{\bra{k} V \ket{0}}\abs{\bra{k} V \ket{\phi}}    
    \end{equation}
    so that $\dist(V\ket{0}, T_{V\ket{\phi}}) = 1 - Z_d^2$ by Lemma \ref{lm:distance-from-torus}. Let $\mu_d = \E Z_d$ and let $\mu := \E\abs{X}\abs{\alpha X +  \beta  Y   }$, where $X,Y$ are independent standard complex Gaussians. By Lemma \ref{lm:gaussian-approximation} we have $\mu_d \to \mu$ uniformly over $\alpha,\beta$, so also $\mu_d^2 \to \mu^2$. Pick $d$ large enough so that $\mu_d^2 < \mu^2 + \vep^2$ (since the convergence rate is $d^{-1/2}$, it suffices that $\vep > c' d^{-1/4}$ for some $c'$). By Lemma \ref{lm:gaussian-average} we have:
    \begin{equation}
        \mu^2 \leq 1 - c\min\{\abs{\beta}^2 ,\beta_0^2\} < 1 - c \cdot C_0^2 \vep^2 = 1 - 4\vep^2
    \end{equation}
    which implies that:
    \begin{equation}
        \mu_d^2 \leq \mu^2 + \vep^2 < 1 - 3\vep^2
    \end{equation}
     We can now estimate:
    \begin{align}
        &\Pp(\dist(V\ket{0}, T_{V\ket{\phi}}) < \vep  ) =
        \Pp( 1 - Z_d^2  < \vep  ) =\\
        & \Pp( Z_d  > \sqrt{1-\vep^2}  ) = \Pp( Z_d  - \mu_d > \sqrt{1-\vep^2} - \mu_d  ) \leq
        \Pp( Z_d  - \mu_d > \sqrt{1-\vep^2} - \sqrt{1-3\vep^2} )
    \end{align}
    By the inequality $\sqrt{1-x} - \sqrt{1-y} \geq \frac{1}{2}(y-x)$ we have $\sqrt{1-\vep^2} - \sqrt{1-3\vep^2} \geq \vep^2$, so we are finally left with bounding:
    \begin{equation}
        \Pp( Z_d  - \mu_d > \vep^2)
    \end{equation}
    We can now apply concentration of measure. By Lemma \ref{lm:f-lipschitz}, the random variable $Z_d$ is Lipschitz with respect to the Hilbert-Schmidt norm with constant $1$. Applying Theorem \ref{th:concentration} we obtain:
    \begin{equation}
        \Pp( Z_d  - \mu_d > \vep^2) \leq \exp(-\frac{\vep^2 (d-2)}{12}) \leq \exp(-\frac{\vep^2 d}{6})
    \end{equation}
    as desired.
\end{proof}

\subsection{Leaving the neighborhood of the initial state}

We now estimate the probability that $U_t\ket{0}$ leaves the ball around $\ket{0}$ for times $t\sim \vep$. Recall that $U_t = V^{\dagger}e^{iDt}V$. We have:
\begin{equation}
    \Pp( d(U_t\ket{0}, \ket{0}) < \vep ) = 
    \Pp( \abs{\bra{0}V^{\dagger}e^{iDt}V\ket{0}}^2 > 1 - \vep^2 ) = 
    \Pp( \abs{ \tr( \dyad{0}{0} V^{\dagger}e^{iDt} V) }^2 > 1 - \vep^2 )
\end{equation}
Let us write $f(V, D) := \abs{ \tr( \dyad{0}{0} V^{\dagger}e^{iDt} V) }^2 $. For fixed $D$, by \eqref{eq:haar-average} we have:
\begin{equation}
    \mu(D) := \E_{V}f(V,D) = \frac{1}{d+1}\left(1 + \frac{\abs{\tr(U_t)}^2}{d}\right)
\end{equation}
We split the probability into two parts:
\begin{align}
    &\Pp(  f(V, D) > 1 - \vep^2 ) =  \\
    &\Pp(  f(V, D) > 1 - \vep^2  \ \cap \ \mu(D) \leq 1-2\vep^2) + \Pp(  f(V, D) > 1 - \vep^2  \cap \mu(D) > 1-2\vep^2) \leq \\
    & \Pp(  f(V, D) - \mu(D) > \vep^2) + \Pp(  f(V, D) > 1 - \vep^2  \cap \mu(D) > 1-2\vep^2) \leq \\
    &\Pp(  f(V, D) - \mu(D) > \vep^2) + \Pp(\mu(D) > 1-2\vep^2)
\end{align}
We bound the first term again using concentration of measure. For every $D$, the function $f(V, D)$ is easily checked to be Lipschitz in the Hilbert-Schmidt norm with respect to $V$ with Lipschitz constant $4$. By Theorem \ref{th:concentration}, the first term is therefore bounded by:
\begin{equation}
    \Pp(  f(V, D) - \mu(D) > \vep^2) \leq \exp(-\frac{\vep^2 (d-2)}{12 \cdot 4^2}) \leq \exp(-\frac{\vep^2 d}{96})
\end{equation}
Bounding the second term is equivalent to bounding $\Pp(\abs{\tr(U_t)}^2 > (1 - 2\vep^2)d^2 - 2\vep^2 d )$. This is estimated in the same way as in Theorem \ref{th:main-gue} and is bounded by $< \delta/2$ for $t > C \vep$ for some constant $C>0$. We have thus obtained that for $d$ large enough we have:
\begin{equation}\label{eq:states-identity}
    \Pp( d(U_t\ket{0}, \ket{0}) < \vep ) < \delta
\end{equation}

\begin{theorem}\label{th:complexity-state}
    Fix $\vep, \delta >0$ and the gateset $\gateset$. Let $H$ be drawn from the $\GUE(d)$ ensemble and assume $\vep > cd^{-1/4}$ for some constant $c$ depending on $\delta$ and also $\vep < \frac{\beta_0}{C_0}$, where $\beta_0, C_0$ are absolute constants as in \cref{th:main-state}. Then for any time $t > C\vep$, with $C$ an absolute constant, we have:
    \begin{equation}
    \Pp(  C_{\vep}(U_t\ket{0}) >  k ) > 1-\delta   
    \end{equation}
    with 
    \begin{equation}
        k = \frac{1}{{\log\abs{\gateset}}}\left( C'\vep^2 d - \log(\frac{1}{\delta}) \right)
    \end{equation} and $C'$ an absolute constant. The same result holds for random basis i.i.d Gaussian ensemble.
\end{theorem}

\begin{proof}
    The proof is virtually the same as the proof of Theorem \ref{th:complexity}. We claim that for $t> C\vep$ and $d$ sufficiently large:
\begin{align}
\Pp\left( U_t \ket{0} \in \bigcup_{\ket{\phi} \in \states{k}}B(\ket{\phi}, \vep) \right) < \delta
\end{align}
Let us divide $\states{k} = \states{k}_{0} \cup \states{k}_{1}$, where $\states{k}_{0} := \{\ket{\phi} \in \states{k} \ \vert \ \ket{\phi} \in B(\ket{0}, C_0 \vep) \}$, where $C_0$ is the constant from Theorem \ref{th:main-state}, and $\states{k}_{1}$ contains the remaining elements. For every $\ket{\phi} \in \states{k}_{0}$ we have $B(\ket{\phi}, \vep) \subseteq B(\ket{0}, (C_0+1)\vep)$, so:
\begin{align}
& \Pp\left( U_t \ket{0} \in \bigcup_{\ket{\phi} \in \states{k}}B(\ket{\phi}, \vep) \right) \leq \Pp\left( U_t \ket{0} \in B(\ket{0}, (C_0 + 1)\vep) \right) + \Pp\left( U_t \ket{0} \in \bigcup_{\ket{\phi} \in \states{k}_{1}}B(\ket{\phi}, \vep) \right) \leq \\
& \Pp\left( U_t \ket{0} \in B(\ket{0}, (C_0 + 1)\vep)\right) + \sum_{\ket{\phi} \in \states{k}_{1}}\Pp\left( U_t \ket{0} \in B(\ket{\phi}, \vep) \right)
\end{align}

For every $\ket{\phi} \in \states{k}_{1}$ we have $d(\ket{\phi},\ket{0}) > C_0 \vep $, so we can invoke Theorem \ref{th:main-state}, to obtain:
\begin{equation}
    \sum_{\ket{\phi} \in \states{k}_{1}}\Pp\left( U_t\ket{0} \in B(\ket{\phi}, \vep) \right) \leq
    \abs{\states{k}} \exp\left( - \frac{\vep^2 d}{6} \right) < \frac{\delta}{2}
\end{equation}
if $k < \frac{C'}{\log\abs{\gateset}} \left(\vep^2 d - \log(\frac{1}{\delta})\right)$ for appropriately chosen constant $C'$. Note that this part of the bound is time-independent.

It remains to bound $\Pp\left( U_t\ket{0} \in B(\ket{0}, (C_0+1)\vep) \right)$. To this end we simply invoke \eqref{eq:states-identity} and plug $\vep \to (C_0+1)\vep, \delta \to 1/2\delta$.

The proof for the random basis i.i.d. Gaussian model proceeds in the same fashion.
\end{proof}

\appendix
\addcontentsline{toc}{section}{Appendices}
\section*{Appendices}

\section{Technical details for GUE}\label{sec:appendixA}

\begin{proof}[Proof of Theorem \ref{th:wigner}]
Let $\rho_d$ be the density of $\mu_{H_d}$. The starting point will the following result on the convergence of densities $\rho_d$ to the semicircle density $\rho_{sc}$ \cite{gotze}: there exist absolute constants $C, \gamma > 0$ such that for $x \in [-2 + \gamma d^{-2/3}, 2 - \gamma d^{-2/3}]$ we have:
\begin{equation}\label{eq:gotze}
\abs{\rho_d(x) - \rho_{sc}(x)} \leq \frac{C}{d (4-x^2)}
\end{equation}
Fix $t>0$ and $\vep > 0$. Let us estimate:
\begin{align}\label{eq:densities}
&\abs{\int_{-\infty}^{\infty} e^{itx}\rho_d(x)dx  -\int_{-\infty}^{\infty} e^{itx}\rho_{sc}(x)dx  } \leq
\int_{-\infty}^{\infty} \abs{\rho_d(x) - \rho_{sc}(x)}dx  \leq \nonumber \\
& \int_{-\infty}^{-2+\gamma d^{-2/3}}\abs{\rho_d(x) - \rho_{sc}(x)}dx + \int_{2-\gamma d^{-2/3}}^{\infty}\abs{\rho_d(x) - \rho_{sc}(x)}dx + \int_{-2+\gamma d^{-2/3}}^{2-\gamma d^{-2/3}}\abs{\rho_d(x) - \rho_{sc}(x)}dx
\end{align}
To bound the first two terms, let us pick $\eta \in (0,2)$ such $\mu_{sc}([-\infty, -2+\eta]) = \mu_{sc}([2-\eta, \infty]) < \frac{1}{8}\vep$ -- since $\mu_{sc}$ has an explicit continuous density supported on $[-2,2]$, such an $\eta$ exists and in fact is readily computed that it is sufficient to take $\eta \sim \varepsilon^{2/3}$. For sufficiently large $d$ we also have $\mu_{H_d}([-\infty, -2+\eta]) = \mu_{H_d}([2-\eta, \infty) < \frac{1}{8}\vep$ (this is the classic convergence of GUE to Wigner semicircle law). In fact, Theorem 1.1 from \cite{gotze} implies that it suffices to take $d >\frac{C}{\varepsilon}$ for some constant $C$. By the triangle inequality this implies that:
\begin{align}
&\int_{-\infty}^{-2+\gamma d^{-2/3}}\abs{\rho_d(x) - \rho_{sc}(x)}dx + \int_{2-\gamma d^{-2/3}}^{\infty}\abs{\rho_d(x) - \rho_{sc}(x)}dx \leq \\
&\mu_{d}([-\infty, -2+\eta]) + \mu_{sc}([-\infty, -2+\eta]) + \mu_{d}([2-\eta, \infty) + \mu_{sc}([2-\eta, \infty) < \frac{1}{2}\vep
\end{align}
We need to take $d$ large enough so that $\gamma d^{-2/3} < \eta$ and since $\eta \sim \varepsilon^{2/3}$, this is again equivalent to $d >\frac{C}{\varepsilon}$. This bounds the first two terms in \eqref{eq:densities}.

To bound the last term, we use \eqref{eq:gotze} to estimate:
\begin{align}
& \int_{-2+\gamma d^{-2/3}}^{2-\gamma d^{-2/3}}\abs{\rho_d(x) - \rho_{sc}(x)}dx \leq \frac{C}{d}\int_{-2+\gamma d^{-2/3}}^{2-\gamma d^{-2/3}}\frac{dx}{ (4-x^2)}dx = \\ \nonumber
&\frac{C}{4d}(\log(4+\gamma d^{-2/3}) + \log(4-\gamma d^{-2/3}) - 2 \log(\gamma d^{-2/3})) = O\left( \frac{\log d}{d}\right) 
\end{align}
which is smaller than $\frac{1}{2}\vep$ for $d$ large enough. Since the first two terms in \eqref{eq:densities} converge as $\frac{1}{d}$ and the last term converges as $\frac{\log d}{d}$, the overall convergence rate is proportional to $\frac{\log d}{d}$.

The claim that $\int_{\R}e^{i\lambda t}d\mu_{sc} (\lambda) = \frac{J_1(2t)}{t}$ is standard computation.
\end{proof}

\begin{theorem}\label{th:hermite}
If $H$ is drawn from the $\GUE(d)$ ensemble, then for any $t$ we have $\Var\tr(U_t) \leq d$.
\end{theorem}

\begin{proof}
We prove the theorem for $H \sim \GUE(d, 1)$, from which the case of $\GUE(d)$ follows by rescaling time by $\sqrt{d}$.

Let $h_n(x)$ be the unnormalized probabilists' Hermite polynomial:
\begin{align}
h_n(x) = (-1)^n e^{x^2/2}\frac{d^n}{dx^n}e^{-x^2/2}
\end{align}
and let $\Psi_n(x)$ be the oscillator wavefunction:
\begin{align}
\Psi_n(x) = \frac{e^{-x^2/4} h_n(x)}{( \sqrt{2\pi}n! )^{1/2}}
\end{align}
Recall that the $1$-point correlation function is a measure with density  $\rho^{(1)}_d(x_1)$ such that for any function $F$ of the eigenvalues we have:
\begin{equation}
    \int F(\lambda)\rho^{(1)}_d(\lambda)d\lambda= \frac{1}{d}\E\sum_{i} F(\lambda)
\end{equation}
and likewise the $2$-point correlation function is a measure with density  $\rho^{(2)}_d(x_1,x_2)$ such that for any function $F$ of the eigenvalues we have:
\begin{equation}
    \int F(\lambda_1, \lambda_2)\rho^{(2)}_d(\lambda_1,\lambda_2)d\lambda_1 d\lambda_2 = \frac{1}{d(d-1)}\E\sum_{i \neq j}F(\lambda_i, \lambda_j)
\end{equation}
We have the following formula for the GUE correlation functions (see \cite{guionnet}, Section 3.2.1):
\begin{align}
\rho^{(1)}_d(x) = \frac{1}{d} K_d(x,x)
\end{align}
and:
\begin{align}
\rho^{(2)}_d(x_1,x_2) = \frac{1}{d(d-1)}\det_{i,j=1}^{2}(K_d(x_i,x_j)) = 
\frac{1}{d(d-1)} \left( K_d(x_1,x_1)K_d(x_2,x_2) - K_d(x_1,x_2)^2 \right)
\end{align}
where $K_d(x,y)$ is the GUE kernel:
\begin{align}
K_d(x,y) = \sum_{k=0}^{d-1}\Psi_k(x)\Psi_k(y)
\end{align}
Since $\E\abs{\Tr(U_t)}^2 = \sum_{k,l}\E e^{i(\lambda_k - \lambda_l)t}$, we have:
\begin{equation}
    \Var\tr(U_t) = \E\abs{\Tr(U_t)}^2 - \abs{\E\Tr(U_t) }^2 = d + \int e^{i(\lambda_1 - \lambda_2)t}\rho^{(2)}_d(\lambda_1, \lambda_2) d\lambda_1 d\lambda_2 - \abs{\E\Tr(U_t) }^2
\end{equation}

Observe that the diagonal term $K_d(\lambda_1, \lambda_1)K_d(\lambda_2, \lambda_2)$ in the integral will exactly reproduce and cancel the squared mean $\abs{\E\Tr(U_t) }^2$. Therefore we only need to handle the off-diagonal term $K_d(\lambda_1,\lambda_2)^2$:
\begin{equation}
    \Var\tr(U_t) = d - \frac{1}{d(d-1)}\int e^{i(\lambda_1 - \lambda_2)t}  K_d(\lambda_1,\lambda_2)^2 d\lambda_1 d\lambda_2 
\end{equation}
It will suffice to prove that the last integral is nonnegative. This follows since we can write:
\begin{align}
K_d(x,y)^2 = \left(\sum_{k=0}^{d-1}\Psi_k(x)\Psi_k(y)\right)^2 = 
\sum_{k,l=0}^{d-1}\Psi_k(x)\Psi_k(y)\Psi_l(x)\Psi_l(y)
\end{align}
which allows us to write the Fourier transform as (switching notation from $\lambda_i$ to $x,y$):
\begin{align}
\int K_d(x,y)^2 e^{i(x-y)t} dx dy = &
\sum_{k,l=0}^{d-1} \int \Psi_k(x)\Psi_k(y)\Psi_l(x)\Psi_l(y) e^{i(x-y)t} dx dy = \\
&\sum_{k,l=0}^{d-1} \abs{\int \Psi_k(x)\Psi_l(x) e^{ixt} dx}^2 
\end{align}
which is manifestly nonnegative.

\end{proof}

\section{Concentration of measure bounds for GUE}\label{sec:appendixB}

The bounds in Section \ref{sec:identity} for GUE are based on moments of $\abs{\tr(U_t)}$ and are therefore only inversely polynomial in $d$. Employing similar strategy as in the case of fixed basis (Section \ref{sec:fixed-basis}), we sketch an exponential bound available for $t = o(d)$ 

Since we are potentially dealing with times $t\sim \vep$, we need a slightly more subtle approach than in Section \ref{sec:fixed-basis}. Observe that for any $\lambda \in [0,1]$ and complex number $z$, if $\abs{z} \geq c$, then at least one of: $\abs{\mathrm{Re}z} \geq \sqrt{1-\lambda}c$ or $\abs{\mathrm{Im}z} \geq \sqrt{\lambda}c$ must hold. By union bound, we thus obtain:
\begin{align}
    &\Pp\left( \abs{\tr(U_t)} > (1-\frac{1}{2}\vep^2) d \right) \leq \\
    &\Pp\left( \abs{\mathrm{Re}\tr (U_t)} > \sqrt{1-\lambda}(1-\frac{1}{2}\vep^2) d \right) + \Pp\left( \abs{\mathrm{Im}\tr(U_t)} > \sqrt{\lambda}(1-\frac{1}{2}\vep^2) d \right)
\end{align}
It is straightforward to check that choosing $\lambda = \vep^2 (1-\vep^2)$ ensures that:
\begin{equation}
    \sqrt{1-\lambda}\left(1-\frac{1}{2}\vep^2 \right) \geq 1- \vep^2
\end{equation}
We are thus left with bounding:
\begin{equation}\label{eq:re}
    \Pp\left( \abs{\mathrm{Re}\tr (U_t)} > (1-\vep^2) d \right)
\end{equation}
and
\begin{equation}\label{eq:im}
    \Pp\left( \abs{\mathrm{Im}\tr(U_t)} > \vep \sqrt{1-\vep^2}\left(1-\frac{1}{2}\vep^2\right) d \right)
\end{equation}
As in Section \ref{sec:fixed-basis}, we first deal with:
\begin{equation}
    \Pp\left( \sum_k \cos(\lambda_k t) > (1-\vep^2) d \right)
\end{equation}
By symmetry of the distribution of GUE eigenvalues $\E\sum_k \cos(\lambda_k t) = d \cdot \E_{\mu_d}e^{it\lambda}$, and for $d$ large enough $\E_{\mu_d}e^{it\lambda}$ can be made smaller than $1-2\vep^2$ for $t > C\vep$ as in the proof of Theorem \ref{th:main-gue}. We are thus left with bounding:
\begin{equation}
    \Pp\left( \frac{1}{d}\left(\sum_k \cos(\lambda_k t) - \E\sum_k \cos(\lambda_k t)\right) > \vep^2  \right) =
    \Pp\left( \frac{1}{d}\left(\tr(\cos(Ht)) - \E\tr(\cos(Ht))\right) > \vep^2  \right)
\end{equation}

We are now in position to use concentration for functions of GUE eigenvalues. We will use the following formulation (follows directly from \cite{guionnet}, Theorem 2.3.5):

\begin{theorem}
Let $f: \R \to \R$ be an $L$-Lipschitz function. If we extend $f$ to a function on symmetric matrices in the usual manner, then for $X \sim \GUE(d, 1/d)$ have:
\begin{align}
\Pp\left( \abs{ \tr(f(X)) - \E\tr(f(X)) } \geq \delta d \right) \leq 2 \exp\left(-\frac{d^2 \delta^2}{4 L^2}\right)
\end{align}
\end{theorem}
In our case $f(x) = \cos(xt)$, which is obviously $t$-Lipschitz, and $\delta = \vep^2$, which leads to:
\begin{align}
   \Pp\left( \frac{1}{d}\left(\sum_k \cos(\lambda_k t) - \E\sum_k \cos(\lambda_k t)\right) > \vep^2  \right) \leq
2 \exp\left(-\frac{d^2 \vep^4}{4t^2 }\right)
\end{align}
which goes to $0$ as $d \to \infty$ if $t = o(d)$. The remaining terms are bounded in the same fashion.

\section{Computation of Gaussian average for states}\label{sec:appendixD}

We proceed to prove Lemma \ref{lm:gaussian-average}. By using the fact that the distribution of $X$ is conjugation invariant we can write:
    \begin{equation}
        \E\abs{X}\abs{\alpha  X +  \beta  Y   } = \E\abs{\alpha  \abs{X}^2 +  \beta X Y   }
    \end{equation}
    Let $\theta$ denote the relative phase of $X$ and $Y$, which is uniformly distributed in $[0,2\pi)$. We then have:
    \begin{equation}
        \E\abs{\alpha  \abs{X}^2 +  \beta X Y   } = \E\abs{\alpha  \abs{X}^2 +  e^{i\theta}\beta \abs{X}\abs{ Y}   }
    \end{equation}
    Without loss of generality we can assume that $\alpha, \beta$ are real. Let us put $x=\abs{X}, y=\abs{Y}$. Since $X=a+ib$, where $a,b$ have variance $1/2$, we have $\abs{X}=\sqrt{a^2+b^2} \sim \frac{1}{\sqrt{2}} \chi_2$, where $\chi_2$ is the chi distribution with density $f(x) = x e^{-x^2/2} \id_{\{x \geq 0\}}$, and likewise for $Y$. This implies that we can write the average over $\abs{X},\abs{Y}$ and $\theta$ as:
    \begin{align}
        &A(\beta) := \E \abs{\alpha  \abs{X}^2 + e^{i\theta} \beta \abs{X}\abs{ Y}} = 
        \E \sqrt{\alpha^2  \abs{X}^4  + \beta^2 \abs{X}^2\abs{ Y}^2 + 2 \alpha\beta\abs{X}^3\abs{Y}\cos\theta }=\\
        & \frac{1}{2}\cdot\frac{1}{2\pi} \int_{0}^{2\pi}d\theta \int_{0}^{\infty}\int_{0}^{\infty}dx dy \sqrt{\alpha^2  x^4  + \beta^2 x^2 y^2 + 2\alpha\beta x^3 y \cos\theta } \cdot x e^{-x^2/2} \cdot y e^{-y^2/2}
    \end{align}
    We handle this integral by passing to polar coordinates $x = r \cos\varphi, y = r \sin\varphi$, where $\varphi \in [0,\pi/2]$ since both $r,s$ are nonnegative.
    \begin{align}
        A(\beta) = \frac{1}{4\pi} \int_{0}^{2\pi}d\theta \int_{0}^{\infty}\int_{0}^{\frac{\pi}{2}}dr d\varphi \cos^2\varphi \sin\varphi \sqrt{\alpha^2 \cos^2\varphi  + \beta^2 \sin^2 \varphi + 2\alpha\beta  \cos\varphi \sin\varphi \cos\theta } \cdot r ^5  e^{-r^2/2}
    \end{align}
    The radial part separates and gives:
    \begin{equation}
        \int_{0}^{\infty} r^5 e^{-r^2 / 2} = 8
    \end{equation}
    so:
    \begin{align}
        &A(\beta) = \frac{2}{\pi}\int_{0}^{2\pi}d\theta \int_{0}^{\frac{\pi}{2}}d\varphi \cos^2\varphi \sin\varphi \sqrt{\alpha^2 \cos^2\varphi  + \beta^2 \sin^2 \varphi + 2\alpha\beta  \cos\varphi \sin\varphi \cos\theta } 
        =\\
        &\frac{2}{\pi}\int_{0}^{2\pi}d\theta \int_{0}^{\frac{\pi}{2}}d\varphi \cos^2\varphi \sin\varphi \sqrt{a^2 + b^2 + 2ab \cos\theta }
    \end{align}
where $a   = \alpha \cos\varphi, b = \beta\sin\varphi $.
    
    We now write:
    \begin{equation}\label{eq:cosines}
        \sqrt{a^2 + b^2 + 2ab \cos\theta } = \sqrt{a^2+b^2}\cdot \sqrt{1 + \frac{2ab}{a^2 + b^2}\cos\theta}
    \end{equation}
    We claim that:
    \begin{equation}\label{eq:square-roots}
        \sqrt{1+x} + \sqrt{1-x} \leq 2 - \frac{1}{4}x^2
    \end{equation}    
    The inequality follows from $\sqrt{1+x} \leq 1 + \frac{1}{2}x - \frac{1}{8}x^2 + \frac{1}{16}x^3$, which follows by repeated differentiation (the terms on the right hand side are the first terms of the Taylor expansion of $\sqrt{1+x}$). Applying \eqref{eq:square-roots} to \eqref{eq:cosines} we arrive at:
    \begin{equation}
        \sqrt{a^2 + b^2 + 2ab \cos\theta } + \sqrt{a^2 + b^2 - 2ab \cos\theta } \leq
        \sqrt{a^2+b^2}\left(2 - \frac{1}{4} \left( \frac{2ab}{a^2 + b^2}\right)^2\cos^2\theta\right)
    \end{equation}
    Since $\cos(x) = -\cos(x+\pi)$ and $\theta$ is uniformly distributed over $[0,2\pi)$, we can apply the above inequality to the integral $A(\beta)$ to obtain:
    \begin{align}
        A(\beta) \leq \frac{2}{\pi}\int_{0}^{2\pi}d\theta \int_{0}^{\frac{\pi}{2}}d\varphi \cos^2\varphi \sin\varphi \sqrt{a^2+b^2}\left(1 - \frac{1}{8} \left( \frac{2ab}{a^2 + b^2}\right)^2\cos^2\theta\right)
    \end{align}
    Substituting back $a=\alpha\cos\varphi, b=\beta\sin\varphi$ and using $\frac{1}{2\pi}\int_{0}^{2\pi}\cos^2 \theta = 1/2$ we can write the inequality as:
    \begin{align}
        A(\beta) \leq I_0(\beta) - \alpha^2 \beta^2 I_1(\beta)
    \end{align}
    where:
    \begin{align}
        &I_0(\beta) = 4  \int_{0}^{\frac{\pi}{2}}d\varphi \cos^2\varphi \sin\varphi \sqrt{\alpha^2\cos^2\varphi +\beta^2\sin^2\varphi}
        \\
&I_1(\beta) = \frac{1}{2} \int_{0}^{\frac{\pi}{2}}d\varphi \cos^2\varphi \sin\varphi \frac{\cos^2\varphi \sin^2 \varphi}{(\alpha^2\cos^2\varphi +\beta^2\sin^2\varphi)^{3/2}}. 
    \end{align}

    One can easily check that the function $I_0$ is nonincreasing, with $I_0(0)=1,I_0(1)=\frac{\pi}{4}$. Moreover, $I_1(\beta)>0$ for all $\beta$ (as for any $\beta$ it is an integral of some nonnegative function that is not identically $0$).  
    
    We now have inserting $\alpha^2 = 1- \beta^2$:
    \begin{equation}
        A(\beta)^2 \leq I_0(\beta)^2 - (1-\beta)^2\beta^2 I_1(\beta)\left( 2 I_0(\beta) - (1-\beta^2)\beta^2I_1(\beta) \right)
    \end{equation}
 Since $I_1>0$, then for $\beta\not=0,1$
$A(\beta)$ is strictly smaller than $I_0(\beta)$, hence the term that is subtracted on right hand side of the above inequality is strictly positive for $\beta\not=0,1$ and it holds also if we drop the factor $(1-\beta^2)\beta^2$:
\begin{align}
    I_1(\beta)\left( 2 I_0(\beta) - (1-\beta^2)\beta^2I_1(\beta) \right)>0,\quad {\rm for} \quad \beta\not=0,1.
\end{align}
But since $I_1$ and $I_2$ are strictly positive for all $\beta$, we obtain that the above expression is strictly positive for all $\beta$. 
Since $I_1$ and $I_2$ are continuous, 
the above expression is therefore bounded from below by some constant $B>0$. 
    Therefore:
    \begin{equation}
        A(\beta)^2 \leq I_0(\beta)^2 - B(1-\beta)^2\beta^2
    \end{equation}
    We easily compute that $I_0(1) = \frac{\pi}{4} < 1$. Choose $\delta  = \frac{1}{2}\left(1-\frac{\pi}{4}\right)$. By continuity there exists a $\beta_0 \in (0,1)$ such that $I_0(\beta) < 1 - \delta$ for $\beta \geq \beta_0$. If $\beta < \beta_0$, then:
    \begin{equation}
        A(\beta)^2 \leq 1 - B(1-\beta_0)^2\beta^2 
    \end{equation}
    and if $\beta \geq \beta_0$:
    \begin{equation}
        A(\beta)^2 \leq 1 - \delta 
    \end{equation}
    If we choose $c := \min\{ B(1-\beta_0)^2 ,\frac{\delta}{\beta_0^2} \}$, we obtain $A(\beta)^2 < 1 - c\min\{\beta^2, \beta_0^2\}$ and the Lemma is proved.

\section{Leaving the identity in the i.i.d. Gaussian model}\label{sec:appendixE}

\begin{proof}[Proof of Proposition \ref{prop:gaussian}]
    
    We can assume that $t < C \vep$ for some constant $C$, since the case $t > C\vep $ is already handled by previous arguments. We first show that $\norm{U_t - I}_{\infty} > \vep$ with high probability -- the case with arbitrary global phase will be similar. Using the independence of $\lambda_k$ we have:
    \begin{align}
        &\Pp\left( \norm{U_t - I}_{\infty} \leq \vep \right) = 
        \Pp\left( \max_{k} \abs{1 - e^{i\lambda_k t}} \leq \vep \right) =\\
        &\Pp\left( \bigcap_{k=1}^{d} \abs{1 - e^{i\lambda_k t}} \leq \vep \right) =
         \Pp\left( \vert\sin(\frac{\lambda t}{2})\vert \leq \frac{\vep}{2} \right)^d
    \end{align}
    Clearly it suffices to show that $ \Pp\left( \vert\sin(\frac{\lambda t}{2})\vert \leq \frac{\vep}{2} \right) < 1 - \frac{a}{d}$ for some constant $a>0$, as $\left( 1 - \frac{a}{d} \right)\to e^{-a}$.

    Clearly $\vert\sin(\frac{\lambda t}{2})\vert \leq \frac{\vep}{2}$ is equivalent to:
    \begin{equation}
        \lambda t \in \bigcup_{l=-\infty}^{\infty}[-\eta + 2l\pi,\eta + 2l\pi]
    \end{equation}
    for some $\eta$ depending on $\vep$. It is easy to see that $c_1 \vep > \eta > c_0 \vep$ for some explicit $c_0, c_1>0$. We can thus union bound:
    \begin{align}
        &\Pp\left( \vert\sin(\frac{\lambda t}{2})\vert \leq \frac{\vep}{2} \right) \leq
        \sum_{l=-\infty}^{\infty}\Pp(\lambda t \in [-\eta + 2l\pi,\eta + 2l\pi]) = \\
        &\Pp(\lambda t \in [-\eta ,\eta ]) + 2 \sum_{l=1}^{\infty}\Pp(\lambda t \in [-\eta + 2l\pi,\eta + 2l\pi])
    \end{align}
    The random variable $\lambda t$ is Gaussian with variance $t^2$, so the tail consisting of terms for $l \geq 1$ can be bounded by:
    \begin{equation}
        \Pp(\lambda t \in [-\eta + 2l\pi,\eta + 2l\pi]) \leq 2\eta \cdot \frac{1}{\sqrt{2\pi} t} \exp(-\frac{(2l\pi - \eta)^2}{2t^2})
    \end{equation}
    Since $\vep < 1/2$, we can take $\eta < \pi$ and estimate the tail by a geometric series:
    \begin{align}
        &2\sum_{l=1}^{\infty}\Pp(\lambda t \in [-\eta + 2l\pi,\eta + 2l\pi]) \leq 2\eta \cdot 2\sum_{l=1}^{\infty}\frac{1}{\sqrt{2\pi} t} \exp(-\frac{(2l\pi - \eta)^2}{2t^2}) \leq\\
        & \frac{4\eta}{\sqrt{2\pi} t} \sum_{l=1}^{\infty} \exp(-\frac{(l\pi)^2}{2t^2}) \leq
        \frac{4\eta}{\sqrt{2\pi} t} \sum_{l=1}^{\infty} \exp(-\frac{l \pi^2}{2t^2}) = 
        \frac{4\eta}{\sqrt{2\pi} t} \cdot \frac{\exp(-\frac{ \pi^2}{2t^2})}{1 - \exp(-\frac{ \pi^2}{2t^2})}
    \end{align}
    The term with $l = 0$ can be estimated using the standard Gaussian tail lower bound:
    \begin{equation}
        \Pp(X > x ) \geq \frac{1}{\sqrt{2\pi}} \cdot \frac{x}{x^2 +1 } \exp(-x^2/2)
    \end{equation}
    we get:
    \begin{equation}
        \Pp(\lambda t \in [-\eta ,\eta ]) = 1 - \Pp(\abs{\lambda t} > \eta ) \leq
        1 - \frac{2}{\sqrt{2\pi}}\frac{\frac{\eta}{t}}{1 + (\frac{\eta}{t})^2} \exp(-\frac{\eta^2}{2t^2})
    \end{equation}
    Since $t < C \vep$ and $c_1 \vep > \eta > c_0 \vep$, we have $\frac{\eta}{t} > A$ for some constant $A$, which implies:
    \begin{equation}
        \Pp(\lambda t \in [-\eta ,\eta ]) \leq 1 - A' \frac{t}{\vep} \cdot \exp(-c_1\frac{\vep^2}{2t^2})
    \end{equation}
    We can assume that $\vep$ is small enough co that $1 - \exp(-\frac{ \pi^2}{2t^2}) > 1/2$. Putting the bounds together leaves us with:
    \begin{align}
                &\Pp\left( \vert\sin(\frac{\lambda t}{2})\vert \leq \frac{\vep}{2} \right) \leq
                1 - A' \frac{t}{\vep} \cdot \exp(-c_1\frac{\vep^2}{2t^2}) + 
                \frac{B\vep}{ t} \cdot \exp(-\frac{ \pi^2}{2t^2})
    \end{align}
    Introduce $y = \frac{\vep}{t}$, so that $C < y < C'\ \sqrt{\log(d)}$ and the bound takes the form:
    \begin{equation}
        1 - A' \frac{1}{y} \cdot \exp(-cy^2) + 
                B y \cdot \exp(-\frac{\pi^2}{2\vep^2} y^2) = 1 - A' \frac{1}{y} \cdot \exp(-cy^2)\left( 1 - \frac{B}{A}y^2\exp(-(\frac{\pi^2}{2\vep^2} - c) y^2)  \right)
    \end{equation}
    By choosing $\vep$ small enough we can make $\frac{\pi^2}{2\vep^2} - c > 0$. The appropriate choice of constants $C, C'$ then ensures that the bound is $< 1 - \frac{a}{d}$ for all $y$ in the chosen range, which finishes the proof. Dealing with arbitrary global phase, i.e. bounding $\norm{U_t - e^{i\varphi}I}_{\infty}$ for arbitrary $\varphi$,  changes only $\sin(\lambda t)$ to $\sin(\lambda t + \varphi)$ in the above expressions and is estimated in virtually the same way.
\end{proof}

\section{Geometric arguments for approximate equidistribtion on the torus}\label{sec:appendix-Haar}

In this Appendix we present a purely geometric argument for the complexity lower bound in the diagonal (fixed basis) case. In contrast to other proofs in this paper, this one is non-probabilistic (does not use concentration of measure or moment bounds) and gives the optimal scaling with respect to $\vep$, namely $\sim d \cdot \log\left(\frac{1}{\vep}\right)$ instead of $\sim d$. 

To explain the intuition behind the proof, we first sketch the setup. Let $H$ be a diagonal random Hamiltonian with eigenvalues $\lambda_1,\dots,\lambda_d$. The time evolution $e^{iHt}$ then happens on the set of diagonal unitary matrices, which we identify with the $d$-dimensional torus $\mathbb{T}^d$. The metric coming from the operator norm is equivalent to the distance on the torus given by:
\begin{equation}\label{eq:torus-metric}
    d((e^{i\varphi_1},\dots, e^{i\varphi_d}), (e^{i\varphi'_1},\dots, e^{i\varphi'_d})  ) = \max_{i=1,\dots,d} \abs{e^{i\varphi_i} - e^{i\varphi'_i}}
\end{equation}

For $t >0$ let $\nu_t$ denote the distribution of $(e^{i\lambda_1 t},\dots, e^{i\lambda_d t})$ and let $\nu$ denote the uniform measure on $\mathbb{T}^{d}$. Now, suppose that the distribution of $\lambda_i$'s has a continuous density $\rho$ (e.g. we exclude the case where $\lambda_i$'s take only countably many values). Then, as $t \to \infty$, the measures $\nu_t$ will converge weakly to $\nu$. Indeed, to check weak convergence on a compact space it suffices to check the convergence of polynomials. In this case this means that for any sequence of natural numbers $\bar{n}=(n_1,\dots,n_d)$ which is not all zeros we must have:
\begin{equation}
    \lim_{t\to\infty}\int_{\mathbb{T}^d} e^{i \scalar{\bar{n}}{\bar{\varphi}}}d\nu_t(\bar{\varphi}) = 
        \int_{\mathbb{T}^d} e^{i \scalar{\bar{n}}{\bar{\varphi}}}d\nu(\bar{\varphi}) = 0
\end{equation}
This follows from the fact that:
\begin{equation}
    \int_{\mathbb{T}^d} e^{i \scalar{\bar{n}}{\bar{\varphi}}}d\nu_t(\bar{\varphi})=
    \int_{\R^d} e^{i \scalar{\bar{n}}{\bar{\lambda}} t}\rho(\bar{\lambda})d\bar{\lambda} 
\end{equation}
and that the density $\rho$ is continuous, so by standard properties of the Fourier transform we have that the Fourier transform of $\rho$ decays to zero at infinity.

The weak convergence of $\nu_t$ to $\nu$ implies that for any measurable set $A \subseteq \mathbb{T}^d$ the measures $\nu_t(A)$ converge to $\nu(A)$. In particular, if we take $A$ to be the $\vep$-ball around any point $x \in \mathbb{T}^d$, for large enough times $t$ the measure $\nu_t(A)$ will be approximately $\sim\vep^d$. This holds for any density $\rho$, but the argument is purely qualitative -- in particular, the convergence time may depend on the dimension $d$. In the case of i.i.d. Gaussian eigenvalues, we show that in fact constant time is sufficient for the measures $\nu_t$ to approximate the uniform measure on $\vep$-balls. 

From now on let $H$ be drawn from the diagonal Gaussian i.i.d. ensemble, with eigenvalues $\lambda_1,\dots,\lambda_d$. We will prove the following Theorem:

\begin{theorem}\label{th:appendix-E-main}
    There exists an absolute constant $C>0$ such that for any $x \in \mathbb{T}^d$, any $t > 1$ and any $\varepsilon$ we have:
\begin{equation}\label{eq:torus-equi-thm}
    \nu_t(B(x, \vep)) \leq C^d \varepsilon^d
\end{equation}
where $B$ denotes the ball in the metric defined in \eqref{eq:torus-metric}.
\end{theorem}

\begin{proof}
It will be convenient to work in the additive model of the torus, namely $\mathbb{T}^{d}_{+} := (\R / 2\pi \Z)^d$, equipped with the metric coming from the quotient of the $\infty$-metric on $\R^d$:
\begin{equation}
d_2(\varphi,\varphi') = \max_{i}\min\{\abs{\varphi_i - \varphi'_i}, \abs{\varphi_i - \varphi'_i \pm 2\pi}\}  
\end{equation}
It is straightforward to prove that:
\begin{equation}
    \frac{2}{\pi}\min\{\abs{\varphi_i - \varphi'_i}, \abs{\varphi_i - \varphi'_i \pm 2\pi}\} \leq \abs{e^{i\varphi_i} - e^{i\varphi'_i}}
\end{equation}
Therefore, we have for any $x \in \mathbb{T}^{d}_{+}$ that $B(x, r) \subseteq B_2(x, \frac{\pi}{2}r)$. Thus, it suffices to prove \eqref{eq:torus-equi-thm} in the metric $d_2$ at the price of a constant $C$ worse by $\frac{\pi}{2}$. By $\nu_t$ we will denote the induced measure both on $\mathbb{T}^{d}$ and $\mathbb{T}^{d}_{+}$.    


Let $\gamma^{(d)}$ denote the i.i.d. Gaussian measure on $\R^d$ and for any $t>0$ let $\gamma^{(d)}_t$ denote the image of $\gamma^{(d)}$ under the map $(\lambda_1, \dots, \lambda_d) \to (\lambda_1 t, \dots, \lambda_d t) $. Note that for any measurable set $S \subseteq \mathbb{T}_{+}$ we have by definition of $\nu_t$:
\begin{equation}\label{eq:torus-equi2}
    \nu_t(S)  = \sum_{y \in (2\pi \Z)^d}\gamma^{(d)}_t(S+y) = \sum_{y \in (2\pi \Z)^d}\gamma^{(d)}\left(\frac{S+y}{t}\right) 
\end{equation}
We now take $S = B_2(x, \varepsilon)$ so that $\frac{S+y}{t} = B_2\left(\frac{x+y}{t}, \frac{\varepsilon}{t}\right)$. We will prove the following statement, which will imply \eqref{eq:torus-equi-thm}:
\begin{equation}\label{eq:torus-equi3}
    \gamma^{(d)}\left(B_2\left(\frac{x+y}{t}, \frac{\varepsilon}{t}\right)\right) \leq
    C^d \varepsilon^d  \gamma^{(d)}\left(B_2\left(\frac{x+y}{t}, \frac{\pi}{t}\right)\right)
\end{equation}
Indeed, by \eqref{eq:torus-equi2}:
\begin{equation}
    \nu_t(B_2\left(x, \varepsilon\right)) = \sum_{y \in (2\pi \Z)^d}\gamma^{(d)}\left( B_2\left(\frac{x+y}{t}, \frac{\varepsilon}{t}\right) \right) \leq
    C^d \varepsilon^d  \sum_{y \in (2\pi \Z)^d}\gamma^{(d)}\left( B_2\left(\frac{x+y}{t}, \frac{\pi}{t}\right) \right) = C^d \varepsilon^d
\end{equation}
where the last equality follows since the balls $B_2\left(\frac{x+y}{t}, \frac{\pi}{t}\right)$ are disjoint and cover all $\R^d$ up to a set of measure of zero. 

Thus, we are reduced to proving the inequality \eqref{eq:torus-equi3}. For intuition behind this inequality, notice that it can be easily rewritten to state that the average of the density over a ball of radius $\frac{\vep}{t}$ should not be much larger than the average over a ball of radius $\frac{\pi}{t}$. We expect this to be true as $t\to\infty$, since at finer and finer scales the density becomes more and more flat. We will now make this intuition quantitative.

We claim that for any $z \in \R^d$, any $\varepsilon < \pi$ and $t > 1$, we have for $C=\frac{1}{\pi} e^{(2\pi+1)^2/2}$:
\begin{equation}
    \gamma^{(d)}\left(B_2\left(z, \frac{\varepsilon}{t}\right)\right) \leq
    C^d \varepsilon^d  \gamma^{(d)}\left(B_2\left(z, \frac{\pi}{t}\right)\right)
\end{equation}
    Since the measure $\gamma^{(d)}$ is a product measure and the ball is a product of balls on each coordinate, it suffices to prove the inequality on each coordinate separately, that is, prove for any $z\in\R$:
    \begin{equation}
        \gamma \left(B\left(z, \frac{\varepsilon}{t}\right)\right) \leq
    C \varepsilon  \gamma \left(B\left(z, \frac{\pi}{t}\right)\right)
    \end{equation}
where $\gamma$ is the $1$-dimensional standard Gaussian measure. This immediately translates into:
\begin{equation}\label{eq:torus-equi4}
    \int_{z - \frac{\varepsilon}{t}}^{z + \frac{\varepsilon}{t}}e^{-x^2/2}dx \leq
    C\varepsilon \int_{z - \frac{\pi}{t}}^{z + \frac{\pi}{t}}e^{-x^2/2}dx
\end{equation}
The function $f(x)=e^{-x^2/2}$ is concave for $\abs{x} < 1$ and convex otherwise. We deal with these possibilities separately, assuming $t > 1$:
\begin{enumerate}
    \item If $\abs{z} < \pi + 1$, then $(z-\pi/t ,z+\pi/t) \subseteq (z-\pi,z+\pi) \subseteq (-2\pi-1,2\pi+1)$, so:
        \begin{equation}
        \int_{z - \frac{\pi}{t}}^{z + \frac{\pi}{t}}e^{-x^2/2}dx \geq \frac{2\pi}{t}e^{-(2\pi+1)^2/2}
    \end{equation}
    and at the same time:
        \begin{equation}
        \int_{z - \frac{\varepsilon}{t}}^{z + \frac{\varepsilon}{t}}e^{-x^2/2}dx \leq \frac{2\varepsilon}{t}
    \end{equation}
    which proves \eqref{eq:torus-equi4} with $C=\frac{1}{\pi} e^{(2\pi+1)^2/2}$
    \item If $\abs{z} \geq \pi + 1$, then the integrand in \eqref{eq:torus-equi4} is convex on the whole range of integration. The desired inequality with $C=\frac{1}{\pi}$ then follows from the fact that if $f$ is convex, then the running average $g(t) := \frac{1}{t}\int_{0}^{t}f(x)dx$ is nondecreasing. To see this, notice that $g'(t) \geq 0$ is equivalent to $\frac{1}{2}(f(0) + f(t) \geq \frac{1}{t}\int_{0}^{t}f(x)dx$, which easily follows from the convexity of $f$ by e.g. approximating the integral by its Riemann sums.
\end{enumerate}
\end{proof}

\begin{proposition}\label{prop:torus-diamond}
    Let $\mathcal{D}$ denote the set of diagonal unitary channels. If $H$ is drawn from the diagonal i.i.d. Gaussian model and $U_t = e^{iHt}$, then for any time $t > 1$ and any $\h{D} \in \mathcal{D}$:
    \begin{equation}
        \Pp(\h{U}_t \in B(\h{D}, \vep)) \leq C^d \vep^{d-1}
    \end{equation}
    where $C$ is an absolute constant and the ball is in the diamond metric.
\end{proposition}

\begin{proof}
By the inequality \eqref{eq:diamond}, it suffices to prove the same statement, but with the diamond metric replaced with the metric $D_{\infty}$. Let $A$ be an $\vep$-net over $[0,2\pi)$ of size $\frac{1}{\vep}$. We clearly have for any choice of representatives $U_t, D$:
\begin{equation}
            \Pp(\h{U}_t \in B_{\infty}(\h{D}, \vep)) = \Pp(U_t \in \bigcup_{\varphi \in A}B(e^{i\varphi}D, \vep)) \leq \sum_{\varphi \in A}\Pp(U_t \in B(e^{i\varphi}D, \vep)) 
\end{equation}
But the probability $\Pp(U_t \in B(e^{i\varphi}D, \vep))$ is simply the measure $\nu_t( B(e^{i\varphi}D, \vep))$ as used before, which by Theorem \ref{th:appendix-E-main} is bounded by $C^d \vep^d$ for $t>1$. The claim of the Proposition follows.
\end{proof}

\begin{theorem}\label{th:complexity-fixed-eps}
        Fix $\vep, \delta >0$ and the gateset $\gateset$. If $H$ is drawn from the diagonal i.i.d. Gaussian model, then for any time $t > 1$ we have:
    \begin{equation}
    \Pp(  C^{\diamond}_{\vep}(\h{U}_t) >  k ) > 1-\delta   
    \end{equation}
    with 
    \begin{equation}
    k = \frac{1}{{\log\abs{\gateset}}} \left(C d \log\left( \frac{1}{C' \vep} \right) - \log(\frac{1}{\delta}) \right)    
    \end{equation}
    and $C, C'$ absolute constants. 
\end{theorem}

\begin{proof}
We start with the union bound:
\begin{equation}\label{eq:union-bound-E}
    \Pp_H(C^{\diamond}_{\vep}(\h{U}_t) \leq k ) = \Pp( \h{U}_t \in \bigcup_{\h{G} \in \gwords{k}} B(\h{G}, \vep ) \leq \sum_{\h{G} \in \gwords{k}} \Pp\left( \h{U}_t \in  B(\h{G}, \vep) \right)
\end{equation}
Let $\mathcal{D}$ be the set of diagonal unitary channels. For a given $\h{G} \in \gwords{k}$, if $\ddiamond(\h{G}, \h{D}) > \vep$ for all $\h{D} \in \mathcal{D}$, then obviously $ \Pp\left( \h{U}_t \in  B(\h{G}, \vep) \right)=0$ as $\h{U}_t$ is always diagonal. Otherwise, there exists some $\h{D} \in \mathcal{D}$ such that $B(\h{G}, \vep) \subseteq B(\h{D},2\vep)$. By Proposition \ref{prop:torus-diamond}, we have thus for some constant $C''$:
\begin{equation}
    \Pp\left( \h{U}_t \in  B(\h{G}, \vep) \right) \leq \Pp\left( \h{U}_t \in  B(\h{D}, 2\vep) \right) \leq (C'')^d 2^{d-1} \vep^{d-1}
\end{equation}
Since there are $\abs{\gateset}^k$ terms in \eqref{eq:union-bound-E}, we have:
\begin{equation}
    \Pp_H(C^{\diamond}_{\vep}(\h{U}_t) \leq k ) \leq \abs{\gateset}^k (C'')^d 2^{d-1} \vep^{d-1} < \delta
\end{equation}
for $k$ as in the statement and appropriate choice of constants $C,C'$.
\end{proof}

\printbibliography
\end{document}